\documentclass{article}

\newcommand{\be}{\begin{equation}}
\newcommand{\ee}{  \end{equation}}
\newcommand{\ba}{\begin{eqnarray}}
\newcommand{\ea}{  \end{eqnarray}}
\newcommand{\ve}{\varepsilon}

\begin{document}

\title{Quantum Graphs and Random-Matrix Theory}

\author{Z. Pluha\v r$^a$ and H. A. Weidenm{\"u}ller$^b$ \\
$^a$Faculty of Mathematics and Physics, Charles University, \\ 180 00 Praha 8, Czech Republic \\ $^b$Max-Planck-Institut f{\"u}r Kernphysik, \\ 69029 Heidelberg, Germany}

\maketitle

\begin{abstract}
For simple connected graphs with incommensurate bond lengths and with
unitary symmetry we prove the Bohigas-Giannoni-Schmit conjecture in
its most general form. Using supersymmetry and taking the limit of
infinite graph size, we show that the generating function for every
$(P, Q)$ correlation function for both closed and open graphs
coincides with the corresponding expression of random-matrix
theory. We use that the classical Perron-Frobenius operator is
bistochastic and possesses a single eigenvalue $+ 1$. In the quantum
case that implies the existence of a zero (or massless) mode of the
effective action. That mode causes universal fluctuation
properties. Avoiding the saddle-point approximation we show that for
graphs that are classically mixing (i.e., for which the spectrum of
the classical Perron-Frobenius operator possesses a finite gap) and
that do not carry a special class of bound states, the zero mode
dominates in the limit of infinite graph size.
\end{abstract}

\section{Introduction}

The distribution of eigenvalues and eigenfuctions of a classically
chaotic Hamiltonian quantum system forms one of the central topics of
quantum chaos. The celebrated conjecture by Bohigas, Giannoni and
Schmit (BGS)~\cite{Boh84} (see also Refs.~\cite{McD79, Cas80, Ber81})
states that {\it the spectral fluctuation properties of a Hamiltonian
  quantum system that is classically chaotic (mixing) coincide with
  those of the random-matrix ensemble in the same symmetry class.}
Here the words ``spectral fluctuation properties'' comprise the
totality of spectral fluctuation measures. The symmetry class
(orthogonal, unitary, or symplectic) is determined~\cite{Dys62} by the
properties of the system under time reversal and under rotation.

Since 1984 the conjecture has found ample numerical support, see
Ref.~\cite{Haa10} and references therein.  However, it took about
twenty years for the first analytical evidence to appear in its
favor. In Refs.~\cite{Sie01, Mue04, Mue05, Heu07, Mue09} the
semiclassical approximation in the form of Gutzwiller's periodic-orbit
theory~\cite{Gut90} was continuously refined to eventually yield a
convincing demonstration of the equality of the level-level correlator
(``two-point function'') of a chaotic (hyperbolic) system and that of
random-matrix theory (RMT), both for unitary and orthogonal
symmetry. The equality holds within an energy interval defined by the
period of the shortest periodic orbit. A parallel effort was devoted
to quantum graphs~\cite{Kot99}. As explained below, these systems,
while not strictly Hamiltonian, are semiclassical from the outset. In
Refs.~\cite{Gnu04, Gnu05} the two-point function for closed quantum
graphs was shown to coincide with that of RMT, both for unitary and
orthogonal symmetry. Open time-reversal-invariant graphs were
considered in Refs.~\cite{Plu13a, Plu13b}. It was shown that the
correlation function of a pair of elements of the scattering matrix
($S$ matrix) is equal to that given by the RMT approach of
Ref.~\cite{Ver85}. As a by-product, the complete distribution function
(given by its moments) of the $S$ matrix in the Ericson regime
(strongly overlapping resonances) was obtained. In the absence of a
corresponding result for RMT it was conjectured that this result is
universal, too.

In this paper we present a proof of the BGS conjecture for quantum
graphs in its most general form. We show that each level correlator
for closed graphs and each S-matrix correlator for open graphs
coincides with the corresponding expression for RMT red in the limit
of infinite graph size.  Not being able to work out these correlators
in general, we demonstrate the equality by showing that their
generating functions are pairwise identical. We define the interval of
wave numbers wherein the equality holds. Since Ref.~\cite{Plu14}
contains a brief account of our results for the orthogonal case, we
focus attention in the present paper on the unitary case. As a
by-product we prove the above-mentioned conjecture concerning Ericson
fluctuations formulated in Refs.~\cite{Plu13a, Plu13b}.

\section{Quantum Graphs}
\label{qua}

A closed graph~\cite{Kot99, Gnu06} is a set of $V$ vertices labeled
$\alpha = 1, \ldots, V$ that are connected by $B$ bonds labeled $b =
1, \ldots, B$. For uniqueness we label the bonds also by the indices
$(\alpha \beta)$ which denote the pair of vertices to which the bond
is attached.  We consider simple connected graphs. In a simple graph,
every pair of non-identical vertices is connected by at most a single
bond, and every bond connects a pair of non-identical vertices. In a
connected graph, starting from any vertex $\alpha$ it is possible to
reach any other vertex $\beta$ through a chain of bonds all of which
belong to the graph. The bond lengths $L_b$ are assumed to be
incommensurate (there exists no linear combination with integer
coefficients $i_b$ such that $\sum_b i_b L_b = 0$). Let $B_\alpha$
denote the number of bonds issuing from vertex $\alpha$.  Then the
total number of bonds is $B = (1/2) \sum_{\alpha = 1}^V B_\alpha$.  We
are interested in generic features and, therefore, consider the limit
$B \to \infty$ of infinite graph size. We assume that in that limit,
the bond lengths remain bounded so that $L_{\rm min} \leq L_b \leq
L_{\rm max}$ with finite $L_{\rm min}, L_{\rm max}$ for all $b$. On
each bond the Schr{\"o}dinger wave has the form $s_{b 1} \exp \{ i k
x_b \} + s_{b 2} \exp \{ - i k x_b \}$ where $x_b$ denotes the
distance to one of the two vertices attached to the bond, and where
the wave number $k$ has the same value on all bonds. The set of
coefficients $\{ s_{b 1}, s_{b 2} \}$ is determined by Hermitean
boundary conditions imposed at each vertex. As a result the vector
${\cal I}^{(\alpha)}$ of incoming waves on all $B_\alpha$ bonds
attached to vertex $\alpha$ and the vector ${\cal O}^{(\alpha)}$ of
outgoing waves on the same bonds are related by ${\cal O}^{(\alpha)} =
\sigma^{(\alpha)} {\cal I}^{(\alpha)}$. The matrices
$\sigma^{(\alpha)}$~\cite{Gnu06} have dimension $B_\alpha$, are
unitary (flux conservation) and, for time-reversal invariant graphs,
are symmetric.

An open graph is obtained from a closed graph as defined in the
previous paragraph by attaching to each one of $\Lambda$ vertices an
additional single bond that extends to infinity. Without loss of
generality these vertices are labeled $\alpha = 1, \ldots,
\Lambda$. The attached additional bonds carry the same labels $\alpha
= 1, \ldots, \Lambda$ and are referred to as channels. We keep the
number of channels $\Lambda$ fixed when we let $B \to
\infty$. Hermitean boundary conditions imposed on all vertices yield
the relations ${\cal O}^{(\alpha)} = \Gamma^{(\alpha)} {\cal
  I}^{(\alpha)}$. Here ${\cal I}^{(\alpha)}$ and ${\cal O}^{(\alpha)}$
are the vectors of incoming and outgoing amplitudes on all the bonds
(channels) attached to vertex $\alpha$. For $\alpha \leq \Lambda$
($\alpha > \Lambda$) these vectors and the matrices
$\Gamma^{(\alpha)}$ have dimension $(B_\alpha + 1)$ ($B_\alpha$,
respectively). The matrices $\Gamma^{(\alpha)}$ are unitary and, for
time-reversal invariant graphs, are symmetric. We write the matrices
$\Gamma^{(\alpha)}$ in the form
\ba
\Gamma^{(\alpha)} &=& \left( \matrix{
             \rho^{(\alpha)} & \tau^{(\alpha)}_\beta \cr
  \tilde{\tau}^{(\alpha)}_\gamma & \sigma^{(\alpha)}_{\gamma \beta} \cr} \right)
 \ {\rm for} \ \alpha \leq \Lambda \ , \nonumber \\
\Gamma^{(\alpha)} &=& (\sigma^{(\alpha)}_{\gamma \beta}) \ \ \ \ {\rm for} \
\alpha > \Lambda \ .
\label{1}
\ea
The coefficient $\rho^{(\alpha)}$ defines the amplitude for
backscattering from channel $\alpha$ into channel $\alpha$. The
coefficient $\tau^{(\alpha)}_\beta$ ($\tilde{\tau}^{(\alpha)}_\gamma)$
defines the amplitude for scattering from bond $(\alpha \beta)$ to
channel $\alpha$ (from channel $\alpha$ to bond $(\alpha \gamma)$,
respectively). The matrices $\sigma^{(\alpha)}$ have dimension
$B_\alpha$, are subunitary (unitary) for $\alpha \leq \Lambda$ (for
$\alpha > \Lambda$, respectively) and, for time-reversal invariant
graphs, are symmetric.

Every set of Hermitean boundary conditions defines a set of unitary
matrices $\{ \sigma^{(\alpha)} \}$ or $\{ \Gamma^{(\alpha)} \}$, as
the case may be. The converse is not neccessarily true. In
constructing the theory we do not specify the boundary conditions but
work with an arbitrary set of unitary matrices $\{ \sigma^{(\alpha)}
\}$ or $\{ \Gamma^{(\alpha)} \}$. That is legitimate: All conclusions
drawn for that set hold also for graphs defined by Hermitean boundary
conditions. 

\subsection{Wave Propagation through a Graph}
\label{wve}

Expressions for the spectral determinant and for the scattering matrix
of graphs have been derived, for instance, in Refs.~\cite{Kot99,
  Gnu06}. For brevity we confine ourselves to a heuristic argument
that highlights the essential points without any claim to rigor. In a
perturbative approach to multiple vertex scattering, the amplitude
$\tilde{{\cal W}}^{- 1}_{\beta \alpha}$ for wave propagation from
vertex $\alpha$ to vertex $\beta$ has the form
\ba
&& \tilde{{\cal W}}^{- 1}_{\beta \alpha} = \exp \{ i k L_{\beta \alpha} \}
\nonumber \\
&& + \sum_\gamma \exp \{ i k L_{\beta \gamma} \} \sigma^{(\gamma)}_{\beta \alpha}
\exp \{ i k L_{\gamma \alpha} \} \nonumber \\
&& + \sum_{\gamma \delta} \exp \{ i k L_{\beta \gamma} \}
\sigma^{(\gamma)}_{\beta \delta} \exp \{ i k L_{\gamma \delta} \}
\sigma^{(\delta)}_{\gamma \alpha} \exp \{ i k L_{\delta \alpha} \} \nonumber \\
&& + \ldots \ .
\label{2}
\ea
To sum that series we introduce matrix notation. The block-diagonal
vertex scattering matrix $\Sigma^{(V)}$ connects incoming and outgoing
amplitudes on all vertices. It carries the $V$ matrices
$\sigma^{(\alpha)}$ with $\alpha = 1, 2, \ldots, V$ in its diagonal
blocks. By definition, $\Sigma^{(V)}$ is unitary for closed and
subunitary for open graphs and has dimension $2 B$, twice the number
$B$ of bonds. Therefore, a doubling of bond indices is indicated. To
that end we introduce ``directed'' bonds. We arrange the $B$ bonds
$(\alpha \beta)$ in lexicographical order so that $\alpha <
\beta$. The resulting series is mapped onto the sequence $b = 1,
\ldots, B$ of integers. The directed bonds in the series are labeled
$(b +)$. To every such directed bond $(\alpha \beta)$ with $\alpha <
\beta$ we associate the bond $(\beta \alpha)$ with opposite direction
and denote it by $(b -)$. With $d = \pm$ the totality of $2 B$
directed bonds is labeled $(b d)$. In directed-bond representation the
matrix $\sigma^{(\alpha)}$ with elements $\sigma^{(\alpha)}_{\beta
  \gamma}$ is written as $\sigma_{\alpha \beta, \alpha \gamma} =
\sigma_{b d, b' d'}$, with the bond labels $b$ ($b'$) determined by
$(\alpha \beta)$ (by $(\alpha \gamma)$, respectively), with $d$
positive (negative) for $\alpha < \beta$ (for $\alpha > \beta$,
respectively), and correspondingly for $d'$. When written in
directed-bond representation, the vertex scattering matrix
$\Sigma^{(V)}$ becomes the ``bond scattering matrix'' $\Sigma^{(B)}$
with elements $\Sigma^{(B)}_{b d, b' d'}$. The map $\Sigma^{(V)} \to
\Sigma^{(B)}$ involves an identical rearrangement of rows and
columns. Therefore, $\Sigma^{(B)}$ is also unitary (subunitary) for
closed (open) graphs, respectively. For time-reversal invariant
graphs, $\Sigma^{(B)}$ is symmetric. The diagonal matrix $\exp \{ i k
{\cal L} \}$ with ${\cal L}_{b d, b' d'} = L_b \delta_{b b'} \delta_{d
  d'}$ describes amplitude propagation on the directed
bonds. Eq.~(\ref{2}) can be summed to give
\be
\tilde{{\cal W}} = \exp \{ - i k {\cal L} \} - \sigma^d_1 \Sigma^{(B)}
\ ,
\label{3}
\ee
with $\sigma^d_1$ the first Pauli spin matrix in directed bond space.
Eq.~(\ref{3}) is verified by expanding $\tilde{{\cal W}}^{- 1}$ in
powers of $\Sigma^{(B)}$. The factor $\sigma^d_1$ is required because
of the definition of $\sigma^{(\alpha)}$ in directed-bond
representation given above. The connection with the definitions used
in Refs.~\cite{Plu13a, Plu13b, Plu14} is established by defining
\be
{\cal W} = \sigma^d_1 \tilde{{\cal W}} = \sigma^d_1 \exp \{ - i k
{\cal L} \} - \Sigma^{(B)} \ .
\label{4}
\ee
We note that $\tilde{\cal W}$ and ${\cal W}$ carry the complete
information on wave propagation through the graph. Therefore, both the
spectral determinant and the scattering matrix can be written in terms
of these matrices. The spectral determinant $\xi(k)$ is~\cite{Gnu04,
  Gnu05}
\ba
\xi(k) &=& \det \{ \exp \{ i k {\cal L} \} \tilde{{\cal W}} \}
\nonumber \\
&=& \det \{ 1 - \exp \{ i k {\cal L} \} \sigma^d_1 \Sigma^{(B)} \} \ .
\label{5}
\ea
Zeros of $\xi(k)$ at $k = k_n$ with $n = 1, 2, \ldots$ define the
bound states of the graph. The level density $d(k)$ of the graph is
given by~\cite{Gnu05}
\be
d(k) = \sum_n \delta(k - k_n) = \langle d_{R} \rangle + d^{\rm fl}(k)
\label{6}
\ee
where
\be
\langle d_R \rangle = \frac{1}{\Delta} = \frac{1}{\pi} \sum_b L_b
\label{7}
\ee
is the average level density and where $\Delta$ denotes the mean level
spacing. The fluctuating part $d^{\rm fl}(k)$ of the level density is
given by~\cite{Gnu05}
\be
d^{\rm fl}(k) = - {1 \over 2 i \pi} \frac{\rm d}{{\rm d} k} \bigg(
\ln \xi(k^+) - \ln \xi(k^-) \bigg) \ .
\label{8}
\ee
Here $k^\pm = k \pm i \epsilon$ with $\epsilon > 0$ and infinitesimal.
Scattering on the graph is described by the scattering matrix $S(k)$,
a function of the wave number $k$. The amplitude $S_{\beta \alpha}(k)$
for scattering from channel $\alpha$ into channel $\beta$ is given
by~\cite{Kot03}
\be
S_{\beta \alpha}(k) = \rho^{(\alpha)} \delta_{\alpha \beta} + \sum_{\gamma \delta}
\tau^{(\beta)}_\gamma \tilde{\cal W}^{- 1}_{\gamma \delta}
\tilde{\tau}^{(\alpha)}_\delta \ .
\label{9}
\ee
In directed-bond representation, the matrices $\tau^{(\beta)}_\gamma$
are written as $\tau_{\beta, \beta \gamma} = \tau_{\beta, b d}$ with
$(b d)$ determined by $(\beta \gamma)$ in terms of the rules stated
above. The totality of these matrices forms the rectangular matrix
${\cal T}$ with $\Lambda$ rows and $2 B$ columns. The matrices
$\tilde{\tau}^{(\alpha)}_\delta$ are similarly written as
$\tilde{\tau}_{\alpha \delta, \alpha} = \tilde{\tau}_{b d,
  \alpha}$. The totality of these matrices forms a rectangular matrix
$\tilde{\cal T}$ with $2 B$ rows and $\Lambda$ columns. With these
definitions the $S$ matrix in Eq.~(\ref{9}) takes the form
\be
S(k) = \rho + {\cal T} {\cal W}^{- 1} \tilde{\cal T} \ .
\label{10}
\ee
Here $\rho$ is diagonal in channel space with elements
$\rho^{(\alpha)}$. The average $S$ matrix $\langle S \rangle$
is given by~\cite{Kot99}
\be
\langle S \rangle = \rho \ .
\label{11}
\ee
The fluctuating part $S^{\rm fl}$ of $S$ is accordingly given by
\be
S^{\rm fl}(k) = {\cal T} {\cal W}^{- 1} \tilde{\cal T} \ .
\label{12}
\ee
For $\alpha \leq \Lambda$ the unitarity of $\Gamma^{(\alpha)}$ in
Eqs.~(\ref{1}) implies
\be
(\sigma^{(\alpha) \dag} \sigma^{(\alpha)})_{\beta \gamma} = \delta_{\beta \gamma} -
\tau^{(\alpha) *}_\beta \tau^{(\alpha)}_\gamma \ .
\label{39}
\ee
The transmission coefficient 
\be
T^{(\alpha)} = \sum_\beta |\tau^{(\alpha)}_\beta|^2 = 1 - |\rho^{(\alpha)}|^2
\label{40}
\ee
measures the unitarity deficit of the average $S$ matrix in
Eq.~(\ref{11}).

\subsection{Chaotic Graphs}
\label{cla}

We consider closed graphs that are chaotic in the classical
limit~\cite{Bar01, Pak01, Gnu06}. In that limit, amplitudes are
replaced by probabilities, and interest centers on the $(2
B)$-dimensional vector $r$ of occupation propabilities $r_{b d} \geq
0$ for the $2 B$ directed bonds $(b d)$. The discrete time evolution
of $r$ is given by the map $r \to {\cal F} r$. Here ${\cal F}$ is the
Perron-Frobenius operator~\cite{Gnu06}, a non-symmetric matrix in
directed-bond space with elements ${\cal F}_{b d, b' d'} = |(\sigma^d_1
\Sigma^{(B)})_{b d, b' d'}|^2$. The classical motion is chaotic
(mixing) if for large times the vector $r$ approaches the uniform
distribution on the set of $2 B$ directed bonds exponentially fast.
That is the case if the spectrum of ${\cal F}$ obeys certain
requirements.

By definition, all elements of ${\cal F}$ are positive or zero.
Moreover, ${\cal F}$ is bistochastic, i.e., $\sum_{b' d'} {\cal F}_{b
  d, b' d'} = 1 = \sum_{b d} {\cal F}_{b d, b' d'}$. This follows from
the unitarity of $\Sigma^{(B)}$ and from the form of $\sigma^d_1$.
For connected graphs, the map $r \to {\cal F} r$ does not possess an
invariant subspace, i.e., the matrix ${\cal F}$ is irreducible. For
matrices with these properties the Perron-Frobenius theorem states
that there exists a non-degenerate maximal eigenvalue $\lambda_1 = 1$
with associated normalized right (left) eigenvectors
\be
u_1 = (1 / \sqrt{2 B}) (1, 1, \ldots, 1)^T \ , \ w_1 = (1 / \sqrt{2 B})
(1, 1, \ldots, 1) \ .
\label{12a}
\ee
All other eigenvalues $\lambda_j$ with $2 \leq j \leq 2 B$ obey
$|\lambda_j| \leq 1$. The associated right (left) eigenvectors $u_j$
($w_j$, respectively) obey $\langle w_j | u_{j'} \rangle = \delta_{j
  j'}$ for $j, j' = 1, \ldots, 2 B$. In general some of the
eigenvalues $\lambda_j$ with $j \geq 2$ may lie on the unit circle in
the complex plane. However, for the graph to be mixing~\cite{Gnu06},
$\lambda_1$ must be the only eigenvalue on that circle. All other
eigenvalues must lie within or on the surface of a disc within the
unit circle. For an $m$-fold repeated map $r \to {\cal F}^m r =
\sum_{j = 1}^{2 B} (\lambda_j)^m u_j \langle w_j | r \rangle$ we then
have $r \to u_1 \langle w_1 | r \rangle$ for $m \to \infty$, and the
uniform distribution is attained exponentially fast. For the graph to
remain mixing in the limit $B \to \infty$ we require that the minimum
distance between the disc of eigenvalues $\lambda_j$ with $j \geq 2$
and the unit circle remains finite, $|\lambda_j| \leq 1 - a$ with $a >
0$ for $B \to \infty$.

For open graphs, the leading eigenvalue differs from unity, and the
leading eigenvector differs from $u_1$ (from $w_1$, respectively).
That difference is taken into account explicitly in our calculation,
see Section~\ref{sad}. We postulate also for open graphs that the
remaining eigenvalues $\lambda_j$ with $j \geq 2$ obey $|\lambda_j|
\leq 1 - a$ with $a > 0$ and that this relation remains valid for $B
\to \infty$.

\subsection{Correlation Functions}
\label{corr}

The fluctuation properties of graphs are completely determined by the
set of all correlation functions. These functions are defined as
averages over $k$ (indicated by angular brackets) and, for closed
graphs, are given by $\langle \prod_{i = 1}^N d^{\rm fl}(k_i) \rangle
$ with $N = 2, 3, \ldots$, taken at arguments $k_1, k_2, \ldots,
k_N$. For open graphs, they are correspondingly given by averages over
products of $N$ fluctuating $S$-matrix elements or their complex
conjugates taken at arguments $k_1, k_2, \ldots, k_N$.  We use the
well-known fact (see also below) that for all $k_1, k_2, \ldots, k_N$
we have
\be
\bigg\langle \prod_{i = 1}^N \frac{\rm d}{{\rm d} k} \ln \xi(k)
\bigg|_{k = k^+_i} \bigg\rangle = 0 = \bigg\langle \prod_{i = 1}^N
S^{\rm fl}(k_i) \bigg\rangle \ .
\label{13}
\ee
Eqs.~(\ref{13}) and (\ref{8}) imply that every correlation function
$\langle \Delta^N \prod_{i = 1}^N d^{\rm fl}(k_i) \rangle $ can be
expressed~\cite{Gnu05} as a linear combination of the dimensionless
$(P, Q)$ level correlation functions defined by
\ba
&& \Delta^{P + Q} \bigg\langle \prod_{p = 1}^P \frac{\rm d}{{\rm d} k}
\ln \xi(k) \bigg|_{k = k^+ + \kappa_p} \prod_{q = 1}^Q \frac{\rm d}{{\rm d}
k} \ln \xi^*(k) \bigg|_{k = k^+ - \tilde{\kappa}_q} \bigg\rangle \nonumber \\
&& = \Delta^{P + Q} \bigg\langle \prod_{p = 1}^P \frac{\rm d}{{\rm d} k}
\ln \xi(k^+ + \kappa_p) \prod_{q = 1}^Q \frac{\rm d}{{\rm d} k} \ln
\xi^*(k^+ - \tilde{\kappa}_q) \bigg\rangle \ .
\label{14}
\ea
Here $P$ and $Q$ are positive integers. Without loss of generality we
take $P \geq Q \geq 1$. In the second line of Eq.~(\ref{14}) we have
used the fact that $k + \kappa_p$ and $k - \tilde{\kappa}_q$ appear
only as arguments of an exponential. We are interested in fluctuations
on the scale of the mean level spacing. Accordingly we require
\be
\langle d_R \rangle \kappa_p, \langle d_R \rangle \tilde{\kappa}_q
\ll B \ .
\label{15}
\ee
This is discussed further in Section~\ref{equ} below. For the
$S$-matrix correlator we have correspondingly
\be
\bigg\langle \prod_{p = 1}^P S^{\rm fl}_{\alpha_p \beta_p}(k +
\kappa_p) \prod_{q = 1}^Q \big(S^{{\rm fl}}_{\gamma_q \delta_q}(k -
\tilde{\kappa}_q)\big)^\dag \bigg\rangle \ ,
\label{16}
\ee
again with $P \geq Q \geq 1$ and with the same bounds~(\ref{15}) on
$\kappa_p$ and $\tilde{\kappa}_q$. The dagger stands for the
combination of transposition and complex conjugation.

The average over wave vector $k$ is carried out over an averaging
interval that is large compared with the minimum difference between
any two bond lengths $L_b$ (interval length $k_0$ with $k_0 \to
\infty$). In calculating the correlation functions, the sequence of
limits is, thus, $\lim_{B \to \infty} \lim_{k_0 \to \infty}$. Because
of the incommensurability of the bond lengths $L_b$ and because of
ergodicity, the average over $k$ is then equivalent to $B$ independent
averages over the phase angles $\phi_b = k L_b$, see
Refs.~\cite{Bar00, Gnu04, Gnu05}. We do not present any details
because this fact is extensively discussed in Refs.~\cite{Gnu04,
  Plu13b}. Eqs.~(\ref{13}) then follow immediately from an expansion
of $\xi(k)$ and of $S^{\rm fl}(k)$ in powers of $\Sigma^{(B)}$. In
what follows, angular brackets always denote phase averages.

We demonstrate the need for incommensurate bond lengths $L_b$ by a
simple example. For $P = 1 = Q$ the two factors in the correlation
functions~(\ref{14}) and (\ref{16}) can each be expanded in powers of
$\Sigma^{(B)}$. The first of these series is proportional to the
right-hand side of Eq.~(\ref{2}) and the second to the Hermitean
adjoint of that expression. We write the product as a joint power
series in $\Sigma^{(B)}$ and $\Sigma^{(B) \dag}$. Each term contains an
exponential. The exponent is a sum of contributions $\pm i k L_b$.
Averaging the term over $k$ gives a nonvanishing result only if these
contributions mutually cancel. If the $L_b$ are incommensurate this is
possible only if all contributions $i k L_b$ appear pairwise with
opposite signs. Under that assumption the color-flavor transformation
in Section~\ref{sup} works and effectively resums the result of the
averaging procedure. Otherwise (i.e., for commensurate $L_b$'s) it
becomes extremely difficult to keep track of all the nonvanishing
contributions, not to speak of resumming the series. But
incommensurability is not a matter of formal convenience only.
Commensurate bond lengths may lead to special states (``topological
resonances''~\cite{Gnu13}) which might affect the spectral statistics.
Thus, incommensurability seems a neccessary condition for graphs to
be chaotic.

\section{Supersymmetry}
\label{sup}

In order to work out the phase averages, we express the correlators in
Eqs.~(\ref{14}) and (\ref{16}) as suitable derivatives of a generating
function ${\cal G}_{\rm G}$. That function is then expressed~\cite{Gnu04,
  Gnu05} as a superintegral. The phase average is worked out with the
help of the color-flavor transformation~\cite{Zir96}.

\subsection{Generating Function}
\label{gen}

For closed graphs we use
\ba
\Delta \frac{\rm d}{{\rm d} k} \ln \xi(k) &=& {\rm Tr} (
\tilde{\cal W}^{- 1} (- i) \Delta {\cal L} \sigma^d_1 \Sigma^{(B)} )
\nonumber \\
&=& \frac{1}{2} \frac{{\rm d}}{{\rm d} j} \frac{\det (\tilde{\cal W}
- i j \Delta {\cal L} \sigma^d_1 \Sigma^{(B)})} {\det( \tilde{\cal W}
+ i j \Delta {\cal L} \sigma^d_1 \Sigma^{(B)})} \bigg|_{j = 0} \ . 
\label{17}
\ea
The differentiation is with respect to the dimensionless source
parameter $j$ at $j = 0$. For open graphs we define in directed-bond
space the matrices
\ba
({\cal S}_p)_{b d, b' d'} &=& \tilde{\cal T}_{b d, \beta_p} {\cal
T}_{\alpha_p, b' d'} \ , \nonumber \\
(\tilde{\cal S}_q)_{b d, b' d'} &=& \tilde{\cal T}_{b d, \gamma_q}
{\cal T}_{\delta_q, b' d'} \ .
\label{18}
\ea
The elements $S^{\rm fl}_{\alpha_p \beta_p}$ and the adjoint elements
$(S^{\rm fl}_{\gamma_q \delta_q})^\dag$ of the fluctuating part of the
$S$ matrix that appear in the correlator~(\ref{16}) can then be
written as
\ba
S^{\rm fl}_{\alpha_p \beta_p} &=& {\rm Tr} ( {\cal W}^{- 1}
{\cal S}_p) = \frac{1}{2} \frac{{\rm d}}{{\rm d} j} \frac{\det
(\tilde{\cal W} + j \sigma^d_1 {\cal S}_p)}{\det( \tilde{\cal W}
- j \sigma^d_1 {\cal S}_p)} \bigg|_{j = 0} \ , \nonumber \\
\big( S^{\rm fl}_{\gamma_q \delta_q} \big)^\dag &=& {\rm Tr} (
{\cal W}^{{- 1} \dag} \tilde{\cal S}^\dag_q) = \frac{1}{2}
\frac{{\rm d}}{{\rm d} j} \frac{\det( \tilde{\cal W}^\dag + j
\tilde{\cal S}^\dag_q \sigma^d_1)} {\det( \tilde{\cal W}^\dag
- j \tilde{\cal S}^\dag_q \sigma^d_1)} \bigg|_{j = 0} \ .
\label{19}
\ea
We define the source terms
\ba
{\cal I}_p &=& - i \Delta {\cal L} \sigma^d_1 \Sigma^{(B)} =
\tilde{\cal I}_q \ {\rm for \ closed \ graphs} \ , \nonumber \\
{\cal I}_p &=& \sigma^d_1 {\cal S}_p \ , \ \tilde{\cal I}_q = \sigma^d_1
\tilde{\cal S}_q \ {\rm for \ open \ graphs} \ .
\label{20}
\ea
In the ratios of determinants in Eqs.~(\ref{17}) and (\ref{19}) we
multiply numerator and denominator both by $\det( \exp \{ i k L \})$
or by the complex conjugate expression. It is convenient to define
\be
z_p = \exp \{ i \kappa_p {\cal L} \} \ , \tilde{z}_q = \exp  \{
i \tilde{\kappa}_q {\cal L} \} \ .
\label{21}
\ee
With these steps we define the generating function ${\cal G}_{\rm G}$
for graphs as
\ba
{\cal G}_{\rm G} &=& \prod_{p = 1}^P \frac{\det( 1 - \exp \{ i k {\cal
L} \} z_p [ \sigma^d_1 \Sigma^{(B)} - j_p {\cal I}_p ])}{\det( 1 -
\exp \{ i k {\cal L} \} z_p [ \sigma^d_1 \Sigma^{(B)} + j_p {\cal
I}_p ])} \nonumber \\
&& \times \prod_{q = 1}^Q \frac{\det( 1 - [ \Sigma^{(B) \dag} \sigma^d_1 -
\tilde{j}_p \tilde{\cal I}^\dag_q ] \tilde{z}_q \exp \{ - i k {\cal L}
\})}{\det( 1 - [ \Sigma^{(B) \dag} \sigma^d_1 + \tilde{j}_q
\tilde{\cal I}^\dag_q ] \tilde{z}_q \exp \{ - i k {\cal L} \})} \ .
\label{22}
\ea
Eqs.~(\ref{17}) and (\ref{19}) then show that the $(P, Q)$ correlation
functions in Eqs.~(\ref{14}) and (\ref{16}) can both be written as
\ba
(P, Q) = \frac{1}{2^{P + Q}} \prod_{p = 1}^P \prod_{q = 1}^Q \frac{{\rm
d}^2}{{\rm d} j_p {\rm d} \tilde{j}_q} {\cal G}_{\rm G} \bigg|_{{\rm all}
\ j = 0} \ .
\label{23}
\ea
Expression~(\ref{22}) allows us to treat open and closed graphs
completely in parallel.

\subsection{Phase Average of the Generating Function}
\label{phas}

With $p = 1, \ldots, P$ and $q = 1, \ldots, Q$ we define the
supervectors $\phi^{(P)}$ with complex commuting elements $s^{(P)}_{p
  b d}$ and anticommuting elements $\chi^{(P)}_{p b d}$ and
$\phi^{(Q)}$ with commuting complex elements $s^{(Q)}_{q b d}$ and
anticommuting elements $\chi^{(Q)}_{q b d}$. These are combined into a
single supervector $\Psi = (\phi^{(P)}, \phi^{(Q)})^T$ of dimension $4
B (P + Q)$. The anticommuting elements obey $\int \chi {\rm d} \chi =
(2 \pi)^{- 1/2} = \int \chi^* {\rm d} \chi^*$. The integration measure
in superspace is the flat Berezinian
\ba
{\rm d} (\tilde{\Psi}, \Psi) &=& \prod_{p = 1}^P \prod_{b d}^{2 B}
{\rm d} \Re (s^{(P)}_{p b d}) {\rm d} \Im (s^{(P)}_{p b d}) {\rm d}
\chi^{* (P)}_{p b d} {\rm d} \chi^{(P)}_{p b d} \nonumber \\
&& \times \prod_{q = 1}^Q \prod_{b' d'}^{2 B} {\rm d} \Re (s^{(Q)}_{q b' d'})
{\rm d} \Im (s^{(Q)}_{q b' d'}) {\rm d} \chi^{* (Q)}_{q b' d'} {\rm d}
\chi^{(Q)}_{q b' d'} \ .
\label{24}
\ea
With $\sigma^s_3$ the third Pauli spin matrix in two-dimensional
Boson-Fermion space, we define the $(4 B)$-dimensional diagonal
supermatrices ${\cal C}_p, \tilde{\cal C}_q, {\cal B}_p$ and
$\tilde{\cal B}_q$ by
\ba
{\cal C}_p &=& \exp \{ i k {\cal L} \} z_p [ \sigma^d_1 \Sigma^{(B)}
- j_p \sigma^s_3 {\cal I}_p ] = \exp \{ i k {\cal L} \} z_p
{\cal B}_p \ , \nonumber \\
\tilde{\cal C}_q &=& \exp \{ i k {\cal L} \} \tilde{z}^*_q [ \sigma^d_1
\Sigma^{(B)} - \tilde{j}_q \sigma^s_3 \tilde{\cal I}_q ] =
\exp \{ i k {\cal L} \} \tilde{z}^*_q \tilde{\cal B}_q \ .
\label{25}
\ea
We define the block-diagonal supermatrix ${\cal C}$ of dimension $4 B
(P + Q)$ that carries the matrices ${\cal C}_p$ ($\tilde{\cal
  C}^\dag_q$) in its first $P$ (in its last $Q$) blocks, respectively,
and analogously for the block-diagonal supermatrix ${\cal B}$. The
projections of ${\cal B}$ onto the retarded (advanced) sector are
denoted by ${\cal B}_+$ (${\cal B}_-$, respectively). With these
definitions the generating function ${\cal G}_{\rm G}$ is written as a
superintegral,
\be
{\cal G}_{\rm G} = \int {\rm d} (\tilde{\Psi}, \Psi) \exp \{ -
\tilde{\Psi} (1 - {\cal C}) \Psi \}
\label{26}
\ee
where $\tilde{\Psi} = (\Psi^*)^T$. The development leading to
Eq.~(\ref{26}) is specific for the case of unitary symmetry. The case
of orthogonal symmetry is slightly more complicated and treated in
Refs.~\cite{Gnu04, Gnu05, Plu13a, Plu13b, Plu14}.

We use Eq.~(\ref{26}) to write the phase average of ${\cal G}_{\rm G}$
as
\ba
\bigg\langle {\cal G}_{\rm G} \bigg\rangle &=& \prod_{b = 1}^B  \int
\frac{ {\rm d} \phi_b}{2 \pi} {\cal G}_{\rm G} \bigg|_{k L_b = \phi_b}
\nonumber \\
&=& \int {\rm d} (\tilde{\Psi}, \Psi) \exp \{ - \tilde{\Psi} \Psi \}
\prod_{b = 1}^B \int \frac{{\rm d} \phi_b}{2 \pi} \exp \{ \tilde{\Psi}
\exp \{ i \Phi \} z {\cal B} \Psi \} \ .
\label{27}
\ea
Here $z$ denotes the diagonal supermatrix with elements $\delta_{b b'}
\delta_{d d'} z_p$ in the retarded and $\delta_{b b'} \delta_{d d'}
z_q$ in the advanced block, jointly referred to as $z_+$ and $z_-$,
respectively. The diagonal supermatrix $\Phi$ has elements $\delta_{d
  d'} \phi_b$ for all $p$ and $- \delta_{d d'} \phi_b$ for all
$q$. The color-flavor transformation~\cite{Zir96} yields
\ba
&& \prod_{b = 1}^B \int \frac{{\rm d} \phi_b}{2 \pi} \exp \{ 
\tilde{\Psi} \exp \{ i \Phi \} z {\cal B} \Psi \} \nonumber \\
&& = \int {\rm d} (\tilde{Z}, Z) {\rm SDet} (1 - Z \tilde{Z}) \exp
\bigg\{ \tilde{\Psi} \left( \matrix{ 0 & Z \cr
                    \tilde{Z} & 0 \cr} \right) z {\cal B} \Psi
\bigg\} \ .
\label{28}
\ea
The matrix $Z$ ($\tilde{Z})$ is rectangular with elements $\delta_{b
  b'} Z_{p b d s; q b d' s'}$ ($\delta_{b b'} \tilde{Z}_{q b d s; p b
  d' s'}$, respectively). The superindex $s = 1, 2$ denotes the
Bosonic and Fermionic degrees of freedom. The Kronecker deltas show
that $Z$ and $\tilde{Z}$ are diagonal in bond space. This fact
reflects the argument formulated at the end of Section~\ref{corr} (all
contributions $i k L_b$ appear pairwise with opposite signs). The
integration measure ${\rm d} (\tilde{Z}, Z)$ is the flat
Berezinian. In Boson-Fermion block notation we have
\be
Z = \left( \matrix{ Z_{B B} & Z_{B F} \cr
                    Z_{F B} & Z_{F F} \cr} \right) \ , \
\tilde{Z} = \left( \matrix{ \tilde{Z}_{B B} & \tilde{Z}_{B F} \cr
                    \tilde{Z}_{F B} & \tilde{Z}_{F F} \cr} \right) \ ,
\label{29}
\ee
with
\be
\tilde{Z}_{B B} = Z^\dag_{B B} \ , \ \tilde{Z}_{F F} = - Z^\dag_{F F} \ .
\label{30}
\ee
Moreover, the eigenvalues of the positive definite Hermitean matrix
$- \tilde{Z}_{F F} Z_{F F}$ are smaller than or equal to unity.

The Gaussian integrals over the variables in $\Psi$ can now be done.
We find
\be
\bigg\langle {\cal G}_{\rm G} \bigg\rangle = \int {\rm d}
(\tilde{Z}, Z) \exp \{ - {\cal A}( \tilde{Z}, Z) \}
\label{31}
\ee
where
\be
{\cal A}(\tilde{Z}, Z) = - {\rm STr} \ln (1 - Z \tilde{Z}) +
{\rm STr} \ln (1 - z_+ {\cal B}_+ Z {\cal B}^\dag_- z_- \tilde{Z} ) \ .
\label{32}
\ee
Here and in what follows the symbol ${\rm STr}$ without (with) indices
denotes the supertrace taken over all matrix indices (only over the
matrix indices indicated). The expression~(\ref{31}) for $\langle
{\cal G}_{\rm G} \rangle$ is exact.

\subsection{Zero Mode}
\label{sad}

We show in Section~\ref{mas} that the effective action ${\cal A}$ in
Eq.~(\ref{32}) contains a massless mode or zero mode. The mode is
defined in terms of the eigenvectors $u_1$ and $w_1$ (see
Eqs.~(\ref{12a})) of the leading eigenvalue $\lambda_1 = 1$ of the
Perron-Frobenius operator as $Y = (Z, u_1), \tilde{Y} = (w_1,
\tilde{Z})$ or, explicitly, as
\ba
Y_{p b d s, q b' d' s'} &=& \delta_{b b'} \delta_{d d'} \frac{1}{\sqrt{2 B}}
\sum_{b'' d''} Z_{p b'' d'' s, q b'' d'' s'} \ , \nonumber \\
\tilde{Y}_{q b d s, p b' d' s'} &=& \delta_{b b'} \delta_{d d'}
\frac{1}{\sqrt{2 B}} \sum_{b'' d''} \tilde{Z}_{q b'' d'' s, p b'' d'' s'} \ .
\label{37}
\ea
The supermatrices $(Y, \tilde{Y})$ are multiples of the unit matrix in
directed bond space. We show in Section~\ref{mas} that under suitable
conditions on the matrix $\Sigma^{(V)}$, the contributions of all
other modes to the correlation functions vanish asymptotically ($B \to
\infty$). Anticipating that result, we express the average generating
functions in terms of the zero-mode contribution only. In that
approximation, $\langle {\cal G}_{\rm G} \rangle$ is
\be
\langle {\cal G}^{(0)}_{\rm G} \rangle = \int \prod_{p q} {\rm d}
(\tilde{Y}_{q p}, Y_{p q}) \exp \{ - {\cal A}^{(0)} \} \ .
\label{40a}
\ee
Here $\prod_{p q} {\rm d} (\tilde{Y}_{q p}, Y_{p q})$ is the flat
Berezinian. The zero-mode contribution to the effective action is
\be
{\cal A}^{(0)} = - {\rm STr} \ln (1 - Y \tilde{Y}) +
{\rm STr} \ln (1 - z_+ {\cal B}_+ Y {\cal B}^\dag_- z_- \tilde{Y} ) \ .
\label{41}
\ee
We use Eqs.~(\ref{21}) and expand ${\cal A}^{(0)}$ in powers of
$\kappa_p$ and $\tilde{\kappa}_q$, putting all $j_p = 0 =
\tilde{j}_q$. Because of the inequalities~(\ref{15}) we keep only
terms up to first order. We use Eq.~(\ref{7}). That yields the
symmetry-breaking term for graphs,
\be
SB_{\rm G} = \frac{i \pi}{\Delta} \sum_p \kappa_p {\rm Str}_s \bigg(
\frac{1}{1 - Y \tilde{Y}} \bigg)_{p p} + \frac{i \pi}{\Delta} \sum_q 
\tilde{\kappa}_q {\rm Str}_s \bigg(\frac{1}{1 - \tilde{Y} Y}
\bigg)_{q q} \ . 
\label{42}
\ee
For open graphs, an additional term arises. It is obtained by putting
in ${\cal A}^{(0)}$ the incremental wave numbers $\kappa_p = 0 =
\tilde{\kappa}_q$ and all source terms $j_p = 0 = \tilde{j}_q$. Since
$\tilde{Y}$ commutes with $\sigma^d_1 \Sigma^{(B)}$ we may use the
cyclic invariance of the trace to write the term bilinear in
$\sigma^d_1 \Sigma^{(B)}$ and $(\sigma^d_1 \Sigma^{(B)})^\dag$ in
Eq.~(\ref{41}) as $Y \Sigma^{(B) \dag} \sigma^d_1 \sigma^d_1
\Sigma^{(B)} \tilde{Y} = Y \Sigma^{(B) \dag} \Sigma^{(B)} \tilde{Y} =
Y \Sigma^{(V) \dag} \Sigma^{(V)} \tilde{Y}$. In the last of these
equations we have switched from directed-bond representation to vertex
representation. That is permissible because $Y$ equals a multiple of
the unit matrix in directed bond space. We take account of the
unitarity deficit of the matrix $\Sigma^{(V)}$ by using
Eqs.~(\ref{39}) and (\ref{40}). We expand the action in powers of $Y$
and $\tilde{Y}$, perform the trace over the bond indices in each term
of the series, and resum the result. Combining that with the first
term on the right-hand side of Eq.~(\ref{41}) we obtain the
channel-coupling term
\be
CC_{\rm G} = - \sum_{\alpha = 1}^\Lambda {\rm STr}_{p s} \ln \bigg( 1 +
T^{(\alpha)} \frac{Y \tilde{Y}}{1 - Y \tilde{Y}} \bigg) \ .
\label{43}
\ee
The average generating function is
\be
\bigg\langle G^{(0)}_{\rm G} \bigg\rangle = \int {\rm d} (\tilde{Y}, Y)
\bigg( \ldots \bigg) \exp \{ SB_{\rm G} + CC_{\rm G} \} \ .
\label{46}
\ee
The dots indicate the source terms (i.e., terms that are linear in
every $j_p$ and in every $\tilde{j}_q$). Only these contribute to
Eq.~(\ref{23}). They are treated in Section~\ref{equ}. For closed
graphs, the channel-coupling term in Eq.~(\ref{46}) is absent.

\subsection{Massive Modes for Graphs}
\label{mas}

\subsubsection{Quadratic Approximation to the Effective Action}

We display the zero mode and the massive modes by expanding the
effective action~(\ref{32}) up to second order in the variables
$\tilde{Z}$, $Z$, putting $z_+ = 1 = z_-$, $\kappa_p = 0 =
\tilde{\kappa}_q$ and dropping the source terms.  That
gives~\cite{Gnu05, Plu13b},
\be
{\cal A}(Z, \tilde{Z}) \approx \sum_{p, q} {\rm STr}_{b d s} \bigg\{
Z_{p q} \tilde{Z}_{q p} - \sigma^d_1 \Sigma^{(B)} Z_{p q} \Sigma^{(B) \dag}
\sigma^d_1 \tilde{Z}_{q p} \bigg\} \ .
\label{33}
\ee
Eq.~(\ref{33}) applies to both closed and open graphs although the
definitions of the matrix $\Sigma^{(B)}$ in the two cases differ. The
matrices $\sigma^d_1$ and $\Sigma^{(B)}$ are the same in all blocks
and do not carry block labels $(p, q)$. The action is a sum over
independent blocks $(p, q)$. We consider a single such contribution
and omit the labels $(p, q)$ for simplicity. In directional space we
decompose
\be
Z = \left( \matrix{ Z_{+ +} & Z_{+ -} \cr
              Z_{- +} & Z_{- -} \cr} \right) = Z_{\rm diag} + Z_{\rm non}
\label{34}
\ee
into diagonal and non-diagonal contributions, and correspondingly for
$\tilde{Z}$. Insertion of this expression into Eq.~(\ref{33})
generates three terms: (i) a term that is bilinear in $Z_{\rm diag}$
and $\tilde{Z}_{\rm diag}$; (ii) a term that is bilinear in $Z_{\rm
  non}$ and $\tilde{Z}_{\rm non}$; (iii) a mixed term.

We first show that the mixed term vanishes~\cite{Gnu05}. Since $Z_{+
  +}, Z_{+ -}, Z_{- +}, Z_{- -}$ are all diagonal in the bond index
$b$, the claim is established if we show that $(\sigma^d_1
\Sigma^{(B)})_{b d_1, b' d'_1} (\Sigma^{(B) \dag} \sigma^d_1)_{b'
  d'_2, b d_2}$ vanishes for $d_1 = d_2$, $d'_1 \neq d'_2$ and for
$d'_1 = d'_2$, $d_1 \neq d_2$. Since $\sigma^d_1$ is off-diagonal in
directional space, that implies $\Sigma^{(B)}_{b d_1, b' d'_1}$
$\times \Sigma^{(B) \dag}_{b' d'_2, b d_2} = 0$ or $\Sigma^{(B)}_{b d_1, b'
  d'_1} \Sigma^{(B) \dag}_{b d_2, b' d'_2} = 0$ for $d_1 = d_2$, $d'_1
\neq d'_2$ and for $d'_1 = d'_2$, $d_1 \neq d_2$.  The argument is the
same for both cases, and we consider $d_1 = d_2$ and $d'_1 \neq
d'_2$. We recall that the matrix $\Sigma^{(B)}$ carries the elements
$\sigma_{\alpha \beta, \alpha \gamma}$ of the matrices
$\sigma^{(\alpha)}$. The pair $(b d_1)$ uniquely determines the pair
$(\alpha \beta)$. The first factor in $\Sigma^{(B)}_{b d_1, b' d'_1}
\Sigma^{(B) *}_{b d_2, b' d'_2}$ vanishes unless the pair $(b', d'_1)$
determines the pair $(\alpha \gamma)$. But then the pair $(b', d'_2)$
with $d'_2 \neq d'_1$ determines the pair $(\gamma \alpha)$ and the
element of the second factor with that index vanishes. This
establishes our claim.

Term (ii) is
\ba
&& \sum_{b b' (d_1 \neq d_2) (d'_1 \neq d'_2)} (\sigma^d_1 \Sigma^{(B)})_{b d_1,
b' d'_1} Z_{b', d'_1 d'_2} (\Sigma^{(B) \dag} \sigma^d_1)_{b' d'_2, b d_2}
\tilde{Z}_{b, d_2 d_1 } \nonumber \\
&& = \sum_{b b' (d_1 \neq d_2) (d'_1 \neq d'_2)} \Sigma^{(B)}_{b (-d_1), b' d'_1}
Z_{b', d'_1 d'_2} \Sigma^{(B) *}_{b (-d_2), b' d'_2} \tilde{Z}_{b, d_2 d_1} \ . 
\label{35}
\ea
For $d_1 \neq d_2$, the pair $(b (-d_1))$ (the pair $(b (-d_2))$)
determines the pair $(\alpha \beta)$ (the pair $(\beta \alpha)$,
respectively). The factor $\Sigma^{(B)}_{b (-d_1), b' d'_1}$ vanishes
unless the pair $(b' d'_1)$ determines a pair $(\alpha \gamma)$. Then
the pair $(b' d'_2)$ with $d'_2 \neq d'_1$ determines the pair
$(\gamma \alpha)$. The result vanishes unless $\beta = \gamma$, i.e.,
unless $(b (-d_1)) = (b' d'_1)$ and $(b (-d_2)) = (b' d'_2)$. This
determines the only nonvanishing contribution. We have $b = b'$ and,
therefore, $-d_1 = d_2$.

Collecting terms we obtain
\ba
&& {\cal A}(Z, \tilde{Z}) \approx {\rm STr}_s \bigg\{
\sum_{b_1 b_2 d_1 d_2} Z_{b_1, d_1 d_1} (1 - {\cal F})_{b_1 d_1, b_2 d_2}
\tilde{Z}_{b_2 , d_2 d_2} \nonumber \\
&& + \sum_{b (d_1 \neq d_2)} Z_{b, d_2 d_1} (1 - \Sigma^{(B)}_{b d_2, b d_2}
\Sigma^{(B) *}_{b d_1, b d_1}) \tilde{Z}_{b, d_1 d_2} \bigg\} \ .
\label{36}
\ea
Here ${\cal F}$ is the Perron-Frobenius operator defined in
Section~\ref{cla}. The last term in Eq.~(\ref{36}) was erroneously
omitted in Ref.~\cite{Gnu05}. To identify the zero mode we first
consider closed graphs. We use the eigenvector decomposition ${\cal
  F}_{b d, b' d'} = \sum_{j = 1}^{2 B} \lambda_j | u_j \rangle \langle
w_j |$ and define, in somewhat symbolic notation, for $j = 1, \ldots,
2 B$ the supermatrices $z_j = \langle Z_{\rm diag} | u_j \rangle$ and
$\tilde{z}_j = \langle w_j | \tilde{Z}_{\rm diag} \rangle$. (We treat
here the matrices $Z_{b d}$ as elements of a linear vector space while
actually they are elements of a coset space. That issue has been
addressed in Refs.~\cite{Zir97, Alt14}). Returning to the full set of
indices, we write Eq.~(\ref{36}) as
\ba
&& {\cal A}(Z, \tilde{Z}) \approx \sum_{p q} \bigg\{ \sum_{j = 2}^{2 B}
(1 - \lambda_j) {\rm STr}_s \bigg( z_{p q, j} \tilde{z}_{q p, j} \bigg)
\nonumber \\
&& + \sum_{b (d_1 \neq d_2)} (1 -
\Sigma^{(B)}_{b d_2, b d_2} \Sigma^{(B) *}_{b d_1, b d_1}) {\rm STr}_s \bigg(
Z_{p b d_2, q b d_1} \tilde{Z}_{q b d_1, p b d_2} \bigg) \bigg\} \ .
\label{38}
\ea
The term with $j = 1$ is conspicuously absent because $\lambda_1 = 1$.
That term defines the zero mode or massless mode $(Y, \tilde{Y})$, see
Eqs.~(\ref{37}). The remaining modes carry ``masses'' $(1 -
\lambda_j)$ and $(1 - \Sigma^{(B)}_{b d_2, b d_2} \Sigma^{(B) *}_{b
  d_1, b d_1})$. The occurrence of the zero mode or massless mode is a
direct consequence of general properties of the Perron-Frobenius
operator ${\cal F}$. The mode exists for all graphs, and it requires
special treatment. The central question is whether $(Y, \tilde{Y})$ is
the only such mode, or whether some of the masses of the other modes
vanish.

Before addressing that question we turn to open graphs. We claim that
Eq.~(\ref{38}) holds also for open graphs. That is obvious for the
last term (even though the values of $\Sigma^{(B)}_{b d_2, b d_2}
\Sigma^{(B) *}_{b d_1, b d_1}$ differ). Some more work is required for
the first term on the right-hand side of Eq.~(\ref{38}). We define the
projector $\Pi_1 = | u_1 \rangle \langle w_1 |$ and the orthogonal
projector $\Pi_\perp = \sum_{j = 2}^{2 B} | u_j \rangle \langle w_j |$
onto two eigenvector spaces of ${\cal F}$ for closed graphs (even
though these are not eigenvector spaces of ${\cal F}$ for open
graphs). We insert $1 = \Pi_1 + \Pi_\perp$ both in front of and right
behind the factor $(1 - {\cal F})$. That gives four terms. The first
term $\Pi_1 (1 - {\cal F}) \Pi_1$ contributes to the channel-coupling
term~(\ref{43}) and is not considered further. Upon diagonalization of
${\cal F}$ the fourth term $\Pi_\perp {\cal F} \Pi_\perp$ takes the
form of the first term on the right-hand side of Eq.~(\ref{38}). Of
the remaining two terms we discuss $\Pi_\perp (1 - {\cal F}) \Pi_1 = -
\Pi_\perp {\cal F} \Pi_1$ (the term $\Pi_1 (1 - {\cal F}) \Pi_\perp$
is treated analogously).  With $u_1 = (1 / \sqrt{2 B}) (1, 1, \ldots,
1)^T$, we have ${\cal F} | u_1 \rangle_{b d} = (1 / \sqrt{2 B})
(\sigma^d_1 \Sigma^{(B)} \Sigma^{(B) \dag} \sigma^d_1)_{b d, b d}$,
and $\Pi_\perp {\cal F} \Pi_1$ is seen to be determined by the
unitarity deficit of $\Sigma^{(B)}$. We switch from directed-bond
representation to vertex representation. We recall that $\Sigma^{(V)}$
is block-diagonal with the matrices $\sigma^{(\alpha)}$ as diagonal
entries. From Eqs.~(\ref{39}) we then have $(\sigma^d_1 \Sigma^{(B)}
\Sigma^{(B) \dag} \sigma^d_1)_{b d, b d} \leftrightarrow (\sigma^d_1
\Sigma^{(B)} \Sigma^{(B) \dag} \sigma^d_1)^{(V)}_{\alpha \beta, \alpha
  \beta} = \sum_\delta |\sigma^{(\beta)}_{\alpha \delta}|^2 = 1 -
|\tau^{(\beta)}_\alpha|^2$. The entire term $- \Pi_\perp {\cal F}
\Pi_1$ is then $\sum_j m_j {\rm STr}_s (\tilde{z}_j Y)$, with the
coupling constant $m_j$ given by $m_j = (1 / \sqrt{2 B}) \sum_{\alpha
  \beta} w_{j, \alpha \beta} |\tau^{(\beta)}_\alpha|^2$. We show that
$m_j \to 0$ for $B \to \infty$ by proving that $\sum_j |m^2_j| \to 0$
for $B \to \infty$. We use Eq.~(\ref{40}) where $T^{(\beta)}$ with $0
\leq T^{(\beta)} \leq 1$ is the transmission coefficient in channel
$\beta$. Using completeness in the form $\sum_{j = 2}^\infty u_{j,
  \alpha \beta} w_{j, \alpha' \beta'} = \delta_{\alpha \alpha'}
\delta_{\beta \beta'} - u_{1, \alpha \beta} w_{1, \alpha' \beta'}$, we
calculate the sum of the $|m^2_j|$ as
\be
\sum_j |m^2_j| = \frac{1}{2 B} \bigg( \sum_{\alpha \beta}
|\tau^{(\beta)}_{\alpha}|^4 - \frac{1}{2 B} (\sum_{\beta} T^{(\beta)})^2 \bigg) \ .
\label{39a}
\ee
For fixed $\beta$ we have $\sum_\alpha |\tau^{(\beta)}_{\alpha}|^4
\leq (T^{(\beta)})^2$. Therefore, the first term on the right-hand
side is positive and bounded by $\Lambda / (2 B)$. The last term is
bounded in magnitude by $\Lambda^2 / (2 B)^2 \ll \Lambda / (2 B)$.
Thus, $\sum_j |m^2_j| \leq \Lambda /(2 B)$. That result is exact and
shows that for $\Lambda$ fixed and $B \to \infty$, all coefficients
$m_j$ vanish. Therefore, the weight factor~(\ref{38}) for the Gaussian
superintegrals over the massive modes holds also for open graphs.

\subsubsection{Loop Expansion}

Generalizing the approach of Refs.~\cite{Gnu04, Gnu05} we show that
the contribution of massive modes to all $(P, Q)$ correlation
functions becomes negligible for $B \to \infty$. We do so by using the
quadratic approximation~(\ref{38}) to the effective action, by
expanding the remaining terms containing massive modes in power
series, and by evaluating the resulting Gaussian superintegrals.

In calculating the Gaussian superintegrals it must be borne in mind
that they differ from ordinary Gaussian integrals. The integration
extends over both commuting and anticommuting integration variables.
We recall that in the Fermion-Fermion sector, the eigenvalues of the
matrices $Z^\dag Z$ are bounded. The same condition is obviously met
by the Fermion-Fermion sector of the transformed variables $(Y,
\tilde{Y})$ and $(z_j, \tilde{z}_j)$ with $j \geq 2$. The actual value
of the bound is immaterial. It can be changed using supersymmetry and
a rescaling of the integration variables. All that matters is that in
the Fermion-Fermion sectors, the masless mode $(Y, \tilde{Y})$ and the
massive modes $(z_j, \tilde{z}_j)$ with $j \geq 2$ require a compact
parametrization. An infinite range of integration occurs only for the
variables in the Boson-Boson blocks.

That last fact defines the conditions under which the Gaussian
superintegrals exist that are obtained from the quadratic
approximation~(\ref{38}) to the effective action: All masses must be
positive. For the series generated in the loop expansion to converge,
that statement must be sharpened. Convergence is assured under the
following two conditions. (i) The eigenvalues $\lambda_j$ with $j \geq
2$ of the Perron-Frobenius operator must lie within or on the surface
of a disk that lies entirely within the unit circle in the complex
plane. That is the same condition as stated in Section~\ref{cla} for
classical mixing of graphs. It must hold for both closed and open
graphs. (ii) In the last term of Eq.~(\ref{38}) that same condition
must be met by all terms $|\Sigma^{(B)}_{b d_2, b d_2} \Sigma^{(B)
  *}_{b d_1, b d_1}|$. Since $\Sigma^{(B)}_{b d, b d} \leftrightarrow
\Sigma^{(V)}_{\alpha \beta, \alpha \beta} = \sigma^{(\alpha)}_{\beta
  \beta}$ we have $\Sigma^{(B)}_{b d_2, b d_2} \Sigma^{(B) *}_{b d_1,
  b d_1} \leftrightarrow \sigma^{(\alpha)}_{\beta \beta}
\sigma^{(\beta)}_{\alpha \alpha}$. We, thus, require that in the limit
$B \to \infty$ we have $|\sigma^{(\alpha)}_{\beta \beta}
\sigma^{(\beta)}_{\alpha \alpha}| \leq b < 1$ for all $\alpha, \beta$.
To interpret that condition let us assume that for some pair of
vertices $(\alpha, \beta)$ we have $|\sigma^{(\alpha)}_{\beta \beta}
\sigma^{(\beta)}_{\alpha \alpha}| = 1$. Unitarity of
$\sigma^{(\alpha)}$ then implies $\sigma^{(\alpha)}_{\beta \gamma} =
0$ and $\sigma^{(\alpha)}_{\gamma \beta} = 0$ for all $\gamma \neq
\beta$, and correspondingly for $\sigma^{(\beta)}$. On the bond
$(\alpha \beta)$ the Schr{\"o}dinger waves are completely
backscattered by both vertices $\alpha$ and $\beta$. The bond is,
thus, completely disconnected from the rest of the graph. It supports
an infinite set of bound states. Condition (ii) excludes the existence
of sets of such states. These would modify the spectral fluctuation
properties of the graph.

The Gaussian superintegrals over the massive modes defined by
Eq.~(\ref{38}) yield unity unless the integrand contains further terms
that depend on $z_j$, $\tilde{z}_j$, $Z_{\rm non}$, $\tilde{Z}_{\rm
  non}$. Such terms are generated by expanding the exponential of the
difference $\delta {\cal A}$ between the effective action in
Eq.~(\ref{32}) and the sum [$SB_{\rm G} + CC_{\rm G}$ plus the
  right-hand side of Eq.~(\ref{38})] in a Taylor series. In the series
we keep only terms that are of first order in every one of the $j_p$
and $\tilde{j}_q$ as only these contribute to the $(P, Q)$ correlation
functions in Eq.~(\ref{23}).

We first consider closed graphs. Then the source terms in
Eq.~(\ref{20}) contain the factor $\Delta {\cal L}$. In the summations
over directed bonds we replace the factors $L_b$ by the average bond
length $\overline{L}$. With $\Delta \overline{L} = \pi / B$, each of
the source terms becomes inversely proportional to $B$. That is
essential for taking the limit $B \to \infty$.  Under omission of the
incremental wave numbers $\kappa_p$, $\tilde{\kappa}_q$ the part
$\delta {\cal A}$ of the action difference $\Delta {\cal A}$ that
contains the source terms is
\be
\delta {\cal A}(Z, \tilde{Z}) = {\rm Str} \ln [ 1 - (1 + i {\bf p}_+)
\sigma^d_1 \Sigma^{(B)} Z (\sigma^d_1 \Sigma^{(B)})^\dag (1 - i {\bf p}_-)
\tilde{Z} ] \ .
\label{46a}
\ee
Here
\be
{\bf p}_+ = (\pi / B) j_+ \sigma^s_3 \ , \ {\bf p}_- = (\pi / B) j_-
\sigma^s_3 \ ,
\label{47}
\ee
and $j_\pm$ are the projections of the source vector $j$ onto the
retarded and the advanced sectors. We note that ${\bf p}_\pm$ and
$\sigma^1_d \Sigma^{(B)}$ commute. The expansion of the exponential
containing $\delta {\cal A}$ generates three types of terms,
\ba
&& {\bf p}_+ (\sigma^d_1 \Sigma^{(B)}) Z (\sigma^d_1 \Sigma^{(B)})^\dag
\tilde{Z} \ , \nonumber \\
&& (\sigma^d_1 \Sigma^{(B)}) Z {\bf p}_- (\sigma^d_1 \Sigma^{(B)})^\dag
\tilde{Z} \ , \nonumber \\
&& {\bf p}_+ (\sigma^d_1 \Sigma^{(B)}) Z {\bf p}_- (\sigma^d_1
\Sigma^{(B)})^\dag \tilde{Z} \ .
\label{48}
\ea
From these three terms the source terms for all $(P, Q)$ correlation
functions are generated.

In the series generated by expanding $\exp \{ \Delta {\cal A} \}$ we
use the transformation that leads from the matrices $Z_{\rm diag},
\tilde{Z}_{\rm diag}$ to the matrices $Y, \tilde{Y}, z_j,
\tilde{z}_j$.  Each of the terms in the series is then a product of
supertraces, each supertrace containing products of the matrices $Y,
\tilde{Y}, z_j, \tilde{z}_j, Z_{\rm non}, \tilde{Z}_{\rm non}$ with
intermittent factors $\sigma^d_1 \Sigma^{(B)}$, its Hermitean adjoint,
source terms, the incremental wave numbers $\kappa_p$ and
$\tilde{\kappa}_q$, the unitary matrix $U$ that diagonalizes ${\cal
  F}$, and its adjoint. The terms that are of order zero in $z_j,
\tilde{z}_j, Z_{\rm non}, \tilde{Z}_{\rm non}$ combine to the source
terms indicated by $( \ldots )$ in Eq.~(\ref{40}) and are discussed in
Section~\ref{equ} below. In the remaining terms we focus attention on
the Gaussian integration over the supermatrices $z_j, \tilde{z}_j,
Z_{\rm non}, \tilde{Z}_{\rm non}$. The integrals obviously vanish
unless in each term of the series the matrices $Y, \tilde{Y}$, $z_j,
\tilde{z}_j$, and $Z_{\rm non}, \tilde{Z}_{\rm non}$ appear in pairs.
Moreover, in every such pair the block indices $(p q)$ on $Y$ (or on
$z_i$ or on $Z_{\rm non}$) must be the same as the block indices $(q
p)$ on $\tilde{Y}$ (or on $\tilde{z}_j$ or on $\tilde{Z}_{\rm non}$,
respectively). The integrals also vanish unless supersymmetry is
broken in both the retarded and the advanced sector of every such pair
by a source term containing the matrix $\sigma^s_3$. When that
condition is not met we say that the integrals vanish because of
supersymmetry. For simplicity we focus attention on terms containing
only the matrices $Y, \tilde{Y}, z_j, \tilde{z}_j$, this being the
slightly more complicated case. Extension of the argument so as to
include the matrices $Z_{\rm non}, \tilde{Z}_{\rm non}$ is completely
straightforward.

We begin with $P = 1 = Q$. In the integration over massive modes the
terms of order one in ${\bf p}_+$ or ${\bf p}_-$ vanish because of
supersymmetry. The only nontrivial contribution to the integrand is
bilinear in ${\bf p}_+$ and ${\bf p}_-$. Under omission of numerical
factors of order unity the terms of lowest order in $Z$, $\tilde{Z}$
are
\ba
&& {\rm STr} \bigg( {\bf p}_+ (\sigma^d_1 \Sigma^{(B)}) Z (\sigma^d_1
\Sigma^{(B)})^\dag \tilde{Z} \bigg) {\rm STr} \bigg( (\sigma^d_1
\Sigma^{(B)}) Z {\bf p}_- (\sigma^d_1 \Sigma^{(B)})^\dag \tilde{Z} \bigg)
\ , \nonumber \\
&& {\rm STr} \bigg( {\bf p}_+ (\sigma^d_1 \Sigma^{(B)}) Z {\bf p}_-
(\sigma^d_1 \Sigma^{(B)})^\dag \tilde{Z} \bigg) \ , \nonumber \\
&& {\rm STr} \bigg( \big[ {\bf p}_+ (\sigma^d_1 \Sigma^{(B)}) Z
(\sigma^d_1 \Sigma^{(B)})^\dag \tilde{Z} \big] \big[ (\sigma^d_1
\Sigma^{(B)}) Z {\bf p}_- (\sigma^d_1 \Sigma^{(B)})^\dag \tilde{Z} \big]
\bigg) \ .
\label{49}
\ea
In the first term we use the transformation leading to Eq.~(\ref{38}),
keep at first only terms of order zero in $(Y, \tilde{Y})$, and obtain
\ba
&& \bigg[ \sum_ {p q} \sum_{j = 2}^{2 B} {\rm STr}_s \bigg\{({\bf p}_+)_p
z_{p q, j} \lambda_j \tilde{z}_{q p, j} \bigg\} \bigg] \nonumber \\
&& \times \bigg[ \sum_ {p' q'} \sum_{j' = 2}^{2 B} {\rm STr}_s \bigg\{
z_{p' q', j} \lambda_{j'} ({\bf p}_-)_{q'} \tilde{z}_{q' p', j'} \bigg\}
\bigg] \ .
\label{50}
\ea
The integral vanishes because of supersymmetry unless $p = p'$, $q =
q'$, $j = j'$. For fixed $(p, q)$ we turn to the remaining single sum
over $j$. Each term in the sum carries a different superintegral. Each
such superintegral is well defined because the integrand is free of
singularities and the infinite range of the bosonic integration
variables is compensated by the Gaussian cutoff. In each of these
integrals we rescale the integration variables so as to remove the
factor $(1 - \lambda_j)$ in the first term on the right-hand side of
Eq.~(\ref{38}). The Berezinian of the matrices $z$ and $\tilde{z}$ is
flat and not affected by the scaling. Therefore, the scaling generates
a factor $1 / (1 - \lambda_j)^2$ multiplying the integral. We also
write the factors $\lambda_j$ and $\lambda_{j'}$ in Eq.~(\ref{50}) in
front of the superintegral. That gives the factor $\lambda^2_j$. Aside
from a relabelling of the integration variables, the remaining
superintegrals are identical for each term in the sum and have the
same finite value ${\cal I}$. The sum takes the form $({\cal I} / B^2)
\sum_{j = 2}^{2 B} \lambda^2_j / (1 - \lambda_j)^2$. That expression
vanishes for $B \to \infty$ if we use $|\lambda_j| \leq a < 1$ for all
$j \geq 2$ as required in Section~\ref{cla}. Avoiding the explicit
calculation of the superintegrals throughout, we use that same method
in the calculation of all the expressions that follow. For brevity we
will say that ``aside from numerical factors'' the Gaussian
superintegrals yield such and such a series.

The second term~(\ref{49}) is
\ba
\sum_{p q} \sum_{j = 2}^{2 B} {\rm Str}_s \bigg\{ ({\bf p}_+)_p z_{p q, j}
\lambda_j ({\bf p}_-)_q \tilde{z}_{q p, j} \bigg\} \ .
\label{51}
\ea
Aside from a numerical factor, the Gaussian superintegrals over this
term with fixed values of $(p, q)$ yield $(1 / B^2) \sum_{j = 2}^{2 B}
\lambda_j / (1 - \lambda_j) \to 0$ for $B \to \infty$. In the third
term~(\ref{49}) we again consider only terms of zeroth order in $Y$
and $\tilde{Y}$. Only those contributions survive where $p = p'$ and
$q = q'$ and where, after the transformation to $z_j$ and
$\tilde{z}_j$, all summation indices $j$ are equal. Except for
numerical factors, the Gaussian integration yields for fixed $(p, q)$
\ba
&& \frac{1}{B^2} \sum_{j = 2}^{2 B} \frac{1}{(1 - \lambda_j)^2}
\sum_{b_1 b_2 b_3 b_4} \sum_{d_1 d_2 d_3 d_4} \bigg\{ (\sigma^d_1
\Sigma^{(B)})_{b_1 d_1, b_2 d_2} w_{b_2 d_2, j} (\sigma^d_1
\Sigma^{(B)})^\dag_{b_2 d_2, b_3 d_3} \nonumber \\
&& \qquad \times  u_{j, b_3 d_3} (\sigma^d_1 \Sigma^{(B)})_{b_3 d_3, b_4 d_4}
w_{b_4 d_4, j} (\sigma^d_1 \Sigma^{(B)})^\dag_{b_4 d_4, b_1 d_1} u_{j, b_1 d_1}
\bigg\} \ .
\label{52}
\ea
Here $u$ and $w$ are the eigenvectors of the PF matrix in
Section~\ref{cla}. Since $\sum_{b' d'} |(\sigma^d_1 \Sigma^{(B)})_{b
  d, b' d'}|^2 = 1$ and $\sum_{b d} u_{j, b d} w_{b d, j} = 1$ for all
$j$, the multiple sum over $b_1, b_2, b_3, b_4$ and $d_1, d_2, d_3,
d_4$ is bounded in magnitude, with a bound common to all values of
$j$, and the expression~(\ref{52}) vanishes for $B \to \infty$.

In addition to the terms~(\ref{49}), the Taylor expansion of $\exp \{
\Delta {\cal A} \}$ generates terms of higher order in $Z$ and
$\tilde{Z}$ that are also linear in $p_+$ and $p_-$. After
transformation to $z_j$ and $\tilde{z}_j$ these vanish unless the
indices $(p, q, j)$ are the same on all $z_j$ and $\tilde{z}_j$. For a
term involving $n$ pairs $z_j, \tilde{z}_j$ the integration yields the
sum $(1 / B^2) \sum_{j = 2}^{2 B} \lambda^m_j / (1 - \lambda_j)^n$
where $m \leq n$. All these sums tend to zero for $B \to \infty$ if
$|\lambda_j| \leq a < 1$ for all $j \geq 2$. As in
expression~(\ref{52}) the remaining factors are products of ordinary
traces over products of factors $\sigma^d_1 \Sigma^{(B)}$, its
Hermitean adjoint, the incremental wave numbers $\kappa_p$ and
$\tilde{\kappa}_q$, and the normalized eigenvectors of ${\cal F}$. All
these traces are bounded from above because $\sigma^d_1 \Sigma^{(B)}$
and $U$ are unitary.  Therefore, the contribution of the massive modes
to the $P = 1 = Q$ correlation function that is of zeroth order in $Y$
and $\tilde{Y}$ vanishes for $B \to \infty$.

In Ref.~\cite{Gnu05} a weaker condition was used to ensure the
vanishing of the contribution due to the massive modes, see also
Ref.~\cite{Tan01}. The convergence of $\sum_{j = 2}^{2 B} \lambda^2_j
/ (1 - \lambda_j)^2$ can be jeopardized only by eigenvalues
$\lambda_j$ close to unity. With eigenvalues ordered such that
$|\lambda_j| \geq\ |\lambda_{j + 1}|$ for all $j$ with $\lambda_1 =
1$, in Ref.~\cite{Gnu05} convergence (and, thereby, vanishing of the
contribution of massive modes) was assured by requesting that for
small $j$ and $B \to \infty$ we have $|\lambda_j| \propto B^{-
  \alpha}$ with $0 \leq \alpha < 1 / 2$. However, a term containing
$n$ pairs $z_j, \tilde{z}_j$ with arbitrary positive integer $n$ as
considered in the previous paragraph (but not considered in
Ref.~\cite{Gnu05}) would impose the stronger bound $\alpha < 1 /
n$. That condition becomes meaningless for $n \to \infty$.

Repeating our arguments for mixed terms containing both the matrices
$(z_j, \tilde{z}_j)$ and the matrices $(Y, \tilde{Y})$ we find that in
all three types of terms that occur for $P = 1 = Q$ such terms do not
arise.

Prior to considering the general case we consider a special term that
arises for $P = 2$, $Q = 1$. With $p_1 \neq p_2$ it is
\ba
&& {\rm STr}_{b d s} \bigg( ({\bf p}_+)_{p_2} (\sigma^d_1 \Sigma^{(B)})
Z_{p_2 1} (\sigma^d_1 \Sigma^{(B)})^\dag \tilde{Z}_{1 p_1} (\sigma^d_1
\Sigma^{(B)}) Z_{p_1 1} ({\bf p}_-)_1 \nonumber \\
&& \qquad \times (\sigma^d_1 \Sigma^{(B)})^\dag \tilde{Z}_{1 p_2} \bigg)
 {\rm STr_{b d s}} \bigg( ({\bf p}_+)_{p_1} (\sigma^d_1 \Sigma^{(B)})
Z_{p_1 1} (\sigma^d_1 \Sigma^{(B)})^\dag \tilde{Z}_{1 p_1} \bigg) \ .
\nonumber \\
\label{53}
\ea
Supersymmetry is broken in the retarded sector in both the $p_1$ and
the $p_2$ blocks by the factors $({\bf p}_+)_{p_1}$ and $({\bf
  p}_+)_{p_2}$.  Supersymmetry in the advanced sector is jointly
broken for both $Z_{p_1, 1}$ and for $\tilde{Z}_{1, p_2}$ by the
single factor $({\bf p}_-)_1$. Therefore, integration over the
term~(\ref{53}) does not yield zero automatically because of
supersymmetry. The example shows that a single symmetry-breaking
matrix $\sigma^s_3$ may be ``shared'' by two different $Z$ matrices.
That insight is important for the general case $P \geq Q \geq 1$.

We use the transformation to $(z, \tilde{z})$, first disregarding $(Y,
\tilde{Y})$. Aside from numerical factors, the term~(\ref{53}) yields
$(1 / B^3) \sum_{j j'} [...]  (1 - \lambda_j)^{- 2} (1 -
\lambda_{j'})^{- 1}$. The factor $[...]$ contains the sums over $(b
d)$ which again are bounded in magnitude, with a bound common to all
values of $j, j'$. Hence, the term~(\ref{53}) vanishes for $B \to
\infty$. A new situation arises when we consider contributions of
higher order in $(z_{p_1 1, j}, \tilde{z}_{1 p_1, j})$ and $(z_{p_2 1,
  j'}, \tilde{z}_{1 p_2, j'})$ with $p_1 \neq p_2$ that share with
expression~(\ref{53}) the property of containing two factors ${\bf
  p}_+$ and one factor ${\bf p}_-$. Such terms arise in the expansion
of $\exp \{ \Delta {\cal A} \}$ and may carry the factors $z_{p_1 1,
  j}, z_{p_2 1, j'}, \tilde{z}_{1 p_1, j}, \tilde{z}_{1 p_2, j'}$ in
intertwined order so that an evaluation of the ensuing Gaussian
superintegrals would not be straightforwardly possible. But the
scaling of all integration variables in the matrices $(z_{p_1 1, j},
\tilde{z}_{1 p_1, j})$ (in $(z_{p_2 2, j'}, \tilde{z}_{2 p_2, j'})$)
with the factor $(1 - \lambda_j)^{1 / 2}$ (with the factor $(1 -
\lambda_{j'})^{1 / 2}$, respectively) works in that case, too,
removing the masses from the Gaussian term in Eq.~(\ref{38}).  In the
integrand, the scaling produces for each value of $j$ a factor $(1 -
\lambda_j)^{- l}$. Here $l$ is the total number of pairs $(z_j,
\tilde{z}_j)$. We then relabel for all values of $j, j'$ the scaled
integration variables $z_j \to z_1$, $z_{j'} \to z_2$, $\tilde{z}_j
\to \tilde{z}_1$, $\tilde{z}_{j'} \to \tilde{z}_2$. The resulting
integrals over $(z_1, \tilde{z}_1)$, $(z_2, \tilde{z}_2)$ are common
to all terms in the sum over $j, j'$, are denoted by ${\cal I}$, and
can be pulled out of the double summation over $j, j'$. (We keep the
notation simple and use here and in what follows the same notation for
the remaining superintegrals although these actually have a different
value in each case.) As a result we obtain $({\cal I} / B^3) \sum_{j
  j'} [...] (1 - \lambda_j)^{- l} (1 - \lambda_{j'})^{- l'}$. The
symbol $[...]$ has the same meaning as before and possesses a bound
common to all values of $j, j'$. The superintegrals in the factor
${\cal I}$ are convergent. As a result, the contribution of the
term~(\ref{53}) vanishes in the limit $B \to \infty$.

In expression~(\ref{53}) the transformation to $(z, \tilde{z})$ may
also produce terms that contain the matrices $(Y, \tilde{Y})$, either
in the form $(Y_{p 1}, \tilde{Y}_{1 p})$ or in the form $(Y_{p 2},
\tilde{Y}_{2 p})$. By way of example we consider the case $(Y_{p 2},
\tilde{Y}_{2 p})$. Omitting the block indices and using the unitarity
of $\sigma^d_1 \Sigma^{(B)}$ we find for the first supertrace in
expression~(\ref{53}) the expression
\be
\sum_j \sum_{b b' d d'} {\rm STr}_{s} \bigg( (p_+) Y (\sigma^d_1
\Sigma^{(B)})^\dag_{b d, b' d'} u_{j, b' d'} \tilde{z}_{j} (\sigma^d_1
\Sigma^{(B)})_{b' d', b d} w_{b d, j} z_{j} (p_-)_1 \tilde{Y} \bigg)
\ .
\label{54}
\ee
Combining that with the second supertrace, we arrive at a single sum
over $j$ only, and the term vanishes for $B \to \infty$ more strongly
than when the matrices $(Y, \tilde{Y})$ are absent.

We turn to the general case. The expansion of $\exp \{ \Delta {\cal A}
\}$ in powers of $(Z, \tilde{Z})$ generates terms of arbitrarily high
orders. We use the transformation to $(z, \tilde{z})$, first
disregarding terms that contain $(Y, \tilde{Y})$. Only terms
containing pairs of matrices $(z_j, \tilde{z}_j)$ with the same index
$j$ and belonging to the same pair of block indices $(p, q)$
contribute to the Gaussian superintegrals. Because of supersymmetry,
the resulting expressions contribute to the massive modes only if they
carry a sufficient number of factors $(p_+)_p$ and $(p_-)_q$. Naively
one might think that for every set of values $(p, q)$ and $j$ that
number is two. However, expression~(\ref{53}) shows that two pairs
$(z_{p q, j}, \tilde{z}_{q p, j})$ and $(z_{p' q', j'}, \tilde{z}_{q'
  p', j'})$ with $j \neq j'$ and $(p, q) \neq (p', q')$ may share a
factor $(p_+)_p$ or a factor $(p_-)_q$, as the case may be. Because of
such sharing, the term with the smallest inverse power of $B$ in the
multiple sum over block indices $(p, q)$ has the form $({\cal I} /
B^{K + 1}) \prod_{n = 1}^K \sum_{j_n = 2}^{2 B} (1 - \lambda)_{j_n}^{-
  l_n} [ \ldots ]$ where the $l_n$ are positive integers. The factor
${\cal I}$ arises when we scale all integration variables in the
manner described above. The factor $[ \ldots ]$ contains products of
sums over products of matrix elements of $S$, of $u$, and of $w$ and
possesses a bound common to all values of $j_n$ for all $n$. The
factor $1 / B^{K + 1}$ accounts for the fact that because of
supersymmetry, at least one pair $(z_j, \tilde{z}_j)$ must carry both
factors $(p_+)_p$ and $(p_-)_q$. Because of the bounds imposed in
Section~\ref{cla} on the eigenvalues $\lambda_j$ for $j \geq 2$ every
sum over one of the $j$'s multiplied with $1 / B$ converges, and the
additional factor $1 / B$ causes the general term to vanish for $B \to
\infty$.

We turn to the terms in the expansion of $\exp \{ \Delta {\cal A} \}$
that are at least linear in $(Y, \tilde{Y})$. Every such term is a
single supertrace or a product of supertraces. In the first case the
matrices $(Y, \tilde{Y})$ appear intertwined with some matrices $(z_j,
\tilde{z}_j)$, and a straightforward generalization of the argument
used for expression~(\ref{54}) shows that the term vanishes more
rapidly than when the $(Y, \tilde{Y})$ are replaced by $(z_j,
\tilde{z}_j)$.  In the case of a product of supertraces some
supertraces may contain only the matrices $(Y, \tilde{Y})$. We then
focus attention on the remaining factors. These may be of order zero
in $(Y, \tilde{Y})$.  Then the arguments of the previous paragraph
apply. Or the remaining factors contain the matrices $(Y, \tilde{Y})$
intertwined with the matrices $(z_j, \tilde{z}_j)$. Then the
straightforward generalization of the argument used for
expression~(\ref{54}) prevails. In all cases every term in the
expansion vanishes for $B \to \infty$.

In conclusion we have shown that for closed graphs, the contribution
of the massive modes $(z_j, \tilde{z}_j)$ vanishes for all $(P, Q)$
correlation functions if the spectrum of the PF operator possesses a
gap of finite size. The reason is that every source term effectively
carries the factor $1 / B$, see Eqs.~(\ref{47}). The same arguments
prevail in the case of the massive modes $(Z_{b d, b -d}, \tilde{Z}_{b
  -d, b d})$ (last term in Eq.~(\ref{38})) provided that
$|\sigma^{(\alpha)}_{\beta \beta} \sigma^{(\beta)}_{\alpha \alpha}|
\leq b < 1$ for all $(\alpha, \beta)$, and in the case of mixed terms
containing both types of massive modes.

To show that the contribution of massive modes vanishes for open
graphs, too, we compare the source terms. According to Eqs.~(\ref{20})
and (\ref{47}) and except for irrelevant factors, the source terms
for closed graphs are given by $(1 /B) \Sigma^{(B)}$ and for open
graphs by ${\cal S}_p$ or $\tilde{\cal S}_q$. According to
Eqs.~(\ref{18}), each matrix ${\cal S}_p$ ($\tilde{\cal S}_q$) is the
dyadic product of a vector ${\cal T}_{\alpha_p}$ and a vector
$\tilde{\cal T}_{\beta_p}$ (of a vector ${\cal T}_{\delta_q}$ and a
vector $\tilde{\cal T}_{\gamma_q}$, respectively). Therefore, only a
single eigenvalue $\rho$ of each of the matrices ${\cal S}_p$ and
$\tilde{\cal S}_q$ differs from zero, with $\rho_p = (T^{(\alpha_p)}
T^{(\beta_p)})^{1/2}$ and $\rho_q = (T^{(\gamma_q)}
T^{(\delta_q)})^{1/2}$ so that $0 \leq \rho \leq 1$ in all cases. In
contradistinction the unitarity of $\Sigma^{(B)}$ for closed graphs
implies that all eigenvalues of that matrix have absolute value
unity. Therefore, traces of matrix products involving ${\cal S}_p$ or
$\tilde{\cal S}_q$ are generically a factor $1 / (2 B)$ smaller than
traces of matrix products involving $\Sigma^{(B)}$. That is the factor
needed to make the contribution of the massive modes disappear also
for open graphs.

\section{Random-Matrix Approach}

We turn to the $(P, Q)$ correlation functions of random-matrix
theory. For systems that are not time-reversal invariant we use the
Gaussian Unitary Ensemble (GUE)~\cite{Meh04}. We define a generalized
generating function ${\cal G}_{\rm R}$ for these functions. We
generalize the supersymmetry approach of Refs.~\cite{Efe83, Ver85,
  Fyo05} and average ${\cal G}_{\rm R}$. We then use the
Hubbard-Stratonovich transformation and the saddle-point
approximation. For closed systems, the CUE~\cite{Zir96} would be an
alternative to the GUE. We use the GUE because we investigate both
closed and open systems, the latter with arbitrarily strong coupling
to the channels.

The complex elements $H_{\mu \nu}$ of the $N$-dimensional Hermitean
GUE Hamiltonian matrix $H$ are Gaussian-distributed random variables
with zero mean values and second moments $\langle H_{\mu \nu}
H^*_{\mu' \nu'} \rangle = (\lambda^2 / N) \delta_{\mu \mu'}
\delta_{\nu \nu'}$. The indices $\mu, \nu$ run from $1$ to $N$. With
$E$ the energy, the $(P, Q)$ level correlation function for the closed
system is defined for integer $P \geq Q \geq 1$ as
\be
d^{(P + Q)} \bigg\langle \prod_{p = 1}^P {\rm Tr} (E^+ + \ve_p - H)^{- 1}
\prod_{q = 1}^Q {\rm Tr} (E^- - \tilde{\ve}_q - H)^{- 1} \bigg\rangle \ .
\label{55}
\ee
Here $E$ denotes the energy and $d$ the mean level spacing. The plus
(minus) sign indicates an infinitesimal positive (negative) imaginary
increment. The angular brackets denote the ensemble average. We are
interested in fluctuations on the scale of the mean level
spacing. Then with $\rho_{\rm R} = 1 / d$ the mean level density, the
energy increments $\ve_p$ and $\tilde{\ve}_q$ obey
\be
\rho_{\rm R} \ve_p \ll \lambda \ , \ \rho_{\rm R} \tilde{\ve}_q \ll
\lambda \ .
\label{56}
\ee

The open system is obtained~\cite{Ver85, Fyo05} by coupling $\Lambda$
channels $a, b, \ldots$ to the states labeled $\mu$ by complex
energy-independent channel-coupling matrix elements $W_{a \mu} =
W^*_{\mu a}$. These obey $(W W^\dag)_{a b} = N v^2_a \delta_{a
  b}$. The unitary scattering matrix is
\be
S_{a b} = \delta_{a b} - 2 \pi i [W (E - H + i \pi W^\dag W)^{- 1}
W^\dag]_{a b} \ .
\label{57}
\ee
As done in Refs.~\cite{Ver85, Fyo05} we have suppressed the shift
matrix. The $(P, Q)$ $S$-matrix correlation function is given by
\be
\bigg\langle \prod_{p = 1}^P S_{a_p b_p}(E + \ve_p)
\prod_{q = 1}^Q \big( S_{c_q d_q}(E - \tilde{\ve}_q) \big)^\dag
\bigg\rangle \ ,
\label{58}
\ee
again with the bounds~(\ref{56}) on $\ve_p$ and $\tilde{\ve}_q$. We
note the strong similarity to the developments in Section~\ref{corr}
but also the following difference. Expression~(\ref{14}) represents
the correlators of the fluctuating parts of the level density only,
while the correlators~(\ref{55}) contain the full level densities,
including the average parts $1 / d$. Since $\Im {\rm Tr} (E^- - H)^{-
  1} = \pi \delta(E - H)$ and with the normalization chosen in
expressions~(\ref{55}) this amounts to the occurence of terms with
value $\pi$ in the final expressions for the correlators. Similarly,
the correlators~(\ref{16}) are defined in terms of the fluctuating
parts of the $S$-matrix elements for graphs while the definitions in
Eqs.~(\ref{61}) and (\ref{63}) below imply that we actually calculate
the RMT correlators of $(S - 1)$ and not of $S$ as suggested by
Eq.~(\ref{58}). That fact yields additional terms $(\langle S \rangle -
1)$ in the final expressions for the RMT correlators. These facts must
be borne in mind when we later compare the source terms for graphs and
for RMT. We calculate the expressions~(\ref{55}) and the correlators
of $(S - 1)$ for $N \to \infty$ in the center of the spectrum where $E
= 0$ and where the mean level spacing $d$ is given by $d = \pi \lambda
/ N$. We use standard conventions: In graph theory (RMT) the mean
level spacing is denoted by $\Delta$ (by $d$) and the mean level
density by $d$ (by $\rho_{\rm R}$, respectively).

\subsection{Generating Function}
\label{genrmt}

To define the generating function we proceed as in Section~\ref{gen}.
For the closed system we have
\be
d {\rm Tr} (E^+ - H)^{- 1} = \frac{1}{2} \frac{\rm d}{{\rm d} j}
\frac{{\rm det}(E^+ - H + d j)}{{\rm det}(E^+ - H - d j)}
\bigg|_{j = 0} \ .
\label{59}
\ee
For the open system we define the $N$-dimensional source matrices
${\cal S}_p$ and $\tilde{\cal S}_q$ with elements
\be
({\cal S}_p)_{\mu \nu} = - 2 i \pi W^*_{\mu a_p} W^{}_{b_p \nu} \ ,
(\tilde{\cal S}_q)_{\mu \nu} = - 2 i \pi W^*_{\mu a_q} W^{}_{b_q \nu} \ .
\label{60}
\ee
Then
\ba
S_{a_p b_p}(E + \ve_p) - \delta_{a b} &=& \frac{1}{2} \frac{\rm d}
{{\rm d} j} \frac{\det(E + \ve_p - H + i \pi W^\dag W + j {\cal S}_p)}
{\det(E + \ve_p - H + i \pi W^\dag W - j {\cal S}_p} \bigg|_{j = 0} \ ,
\nonumber \\
\big( S_{c_q d_q}(E - \tilde{\ve}_q) \big)^\dag - \delta_{a b} &=&
\frac{1}{2} \frac{\rm d}{{\rm d} j} \frac{\det(E - \tilde{\ve}_q -
H - i \pi W^\dag W + j \tilde{\cal S}^\dag_q)}{\det(E - \tilde{\ve}_q
- H - i \pi W^\dag W - j \tilde{\cal S}^\dag_p}) \bigg|_{j = 0} \ .
\nonumber \\
\label{61}
\ea
We define the source terms
\ba
{\cal I}_p &=& d = \tilde{\cal I}_q \ {\rm for \ the \ closed \ system}
\ , \nonumber \\
{\cal I}_p &=& {\cal S}_p \ , \ \tilde{\cal I}_q = \tilde{\cal S}_q
\ {\rm for \ the \ open \ system} \ ,
\label{62}
\ea
and the generating function $G_{\rm R}$ for the random-matrix approach
as
\ba
{\cal G}_{\rm R} &=& \prod_{p = 1}^P \frac{{\rm det}(E^+ + \ve_p - H
+ i \pi \delta_{\rm open} W^\dag W + j_p {\cal I}_p)}{{\rm det}(E^+
+ \ve_p - H + i \pi \delta_{\rm open} W^\dag W - j_p {\cal I}_p)}
\nonumber \\ 
&& \times \prod_{q = 1}^Q \frac{{\rm det}(E^- - \tilde{\ve}_q - H -
i \pi \delta_{\rm open} W^\dag W + \tilde{j}_q \tilde{\cal I}_q)}
{{\rm det}(E^- - \tilde{\ve}_q - H - i \pi \delta_{\rm open} W^\dag W
- \tilde{j}_q \tilde{\cal I}_q)} \ .
\label{63}
\ea
The factor $\delta_{\rm open}$ equals zero (one) for the closed (the
open) system, respectively. With these definitions the $(P, Q)$
correlation functions in Eqs.~(\ref{55}) and (\ref{58}) can both be
written as
\be
(P, Q) = \frac{1}{2^{P + Q}} \prod_{p = 1}^P \prod_{q = 1}^Q \frac{{\rm
d}^2}{{\rm d} j_p {\rm d} \tilde{j}_q} {\cal G}_{\rm R} \bigg|_{{\rm all}
\ j = 0} \ .
\label{64}
\ee

\subsection{Supersymmetry}

With $p = 1, \ldots, P$ and $q = 1, \ldots, Q$ we define the
supervectors $\phi^{(P)}$ with complex commuting elements $s^{(P)}_{p
  \mu}$ and anticommuting elements $\chi^{(P)}_{p \mu}$ and
$\phi^{(Q)}$ with commuting complex elements $s^{(Q)}_{q \mu}$ and
anticommuting elements $\chi^{(Q)}_{q \mu}$. These are combined into a
single supervector $\Psi = (\phi^{(P)}, \phi^{(Q)})^T$ of dimension $2
N (P + Q)$. The anticommuting elements obey $\int \chi {\rm d} \chi =
(2 \pi)^{- 1/2} = \int \chi^* {\rm d} \chi^*$. The integration measure
in superspace is the flat Berezinian
\ba
{\rm d} (\tilde{\Psi}, \Psi) &=& \prod_{p = 1}^P \prod_{\mu = 1}^{N}
{\rm d} \Re (s^{(P)}_{p \mu}) {\rm d} \Im (s^{(P)}_{p \mu}) {\rm d}
\chi^{* (P)}_{p \mu} {\rm d} \chi^{(P)}_{p \mu} \nonumber \\
&& \times \prod_{q = 1}^Q \prod_{\nu = 1}^{N} {\rm d} \Re (s^{(Q)}_{q \nu})
{\rm d} \Im (s^{(Q)}_{q \nu}) {\rm d} \chi^{* (Q)}_{q \nu} {\rm d}
\chi^{(Q)}_{q \nu} \ .
\label{65}
\ea
The index $s = 1, 2$ runs over superspace, the index $t = 1, 2,
\ldots, P + Q$ denotes the retarded $(t \leq P)$ and advanced ($t >
P$) blocks. The generating function can be written as
\be
G_{\rm R} = \int {\rm d} (\tilde{\Psi}, \Psi) \exp \{ \frac{i}{2}
\tilde{\Psi} {\bf L}^{1/2} {\bf D} {\bf L}^{1/2} \Psi \}
\label{66}
\ee
where $\tilde{\Psi} = (\Psi^*)^T$. Here ${\bf L}$ is the third Pauli
spin matrix in retarded-advanced space. The matrix ${\bf D}$ is block
diagonal and given by
\be
{\bf D} = {\bf E} - {\bf H} + \ve + i {\bf W} + {\bf J} \ .
\label{67}
\ee
We define $\ve_t = \ve_p$ for $t \leq P$ and $\ve_t = \tilde{\ve}_q$
for $t = P + q$ and correspondingly for $j_t$. We write ${\cal I}_t =
{\cal I}_p$ for $t \leq P$ and ${\cal I}_t = {\cal I}^\dag_q$
for $t = P + q$. Then
\ba
&& {\bf E} = \{ \delta_{t t'} \delta_{\mu \mu'} \delta_{s s'} E \},
\ {\bf H} = \{ \delta_{t t'} \delta_{s s'} H_{\mu \mu'} \}, \
\ve = \{ \delta_{\mu \mu'} \delta_{s s'} L_{t t'} \ve_t \},
\nonumber \\
&& {\bf W} = \{ \delta_{\rm open} \delta_{s s'} \pi L_{t t'} 
(W^\dag W)_{\mu \mu'} \},
\ {\bf J} = \{ \delta_{t t'} \delta_{\mu \mu'} \sigma^s_3 j_t {\cal I}_t
\} \ .
\label{68}
\ea
We average ${\cal G}_{\rm R}$ over the ensemble by averaging $\exp \{
- (i/2) (\tilde{\Psi} {\bf L}^{1/2} {\bf H} {\bf L}^{1/2} \Psi)
\}$. We have
\be
\bigg\langle \exp \bigg\{ - (i/2) (\tilde{\Psi} {\bf L}^{1/2} {\bf H}
{\bf L}^{1/2} \Psi) \bigg\} \bigg\rangle = \exp \bigg\{ {1 \over 8 N}
{\rm STr} (A^2) \bigg\}
\label{69}
\ee
where
\be
A_{t s, t' s'} = i \lambda ({\bf L}^{1/2})_{t t} \sum_\mu \Psi_{t \mu s}
\Psi^*_{t' \mu s'} ({\bf L}^{1/2})_{t' t'} \ .
\label{70}
\ee
We insert the result of Eqs.~(\ref{69}, \ref{70}) into the expression
for $\langle {\cal G}_{\rm R} \rangle $ and remove the terms that are
quartic in the integration variables by the Hubbard-Stratonovich
transformation. The remaining Gaussian integrals over $\Psi$ and
$\tilde{\Psi}$ can be done. All these steps are standard. With ${\bf
  \Sigma} = \{ \delta_{\mu \mu'} \sigma_{t s, t' s'} \}$ the result is
\be
\bigg\langle {\cal G}_{\rm R} \bigg\rangle = \int {\rm d} [\sigma]
\exp \bigg\{ - \frac{N}{2} {\rm STr}_{t s} (\sigma^2) - {\rm STr} \ln
\bigg( {\bf E} - \lambda {\bf \Sigma} + \ve + i {\bf W} + {\bf J}
\bigg) \bigg\} \ . 
\label{71}
\ee
The matrix $\sigma$ has the same dimension and the same symmetries as
the matrix $A$ in Eq.~(\ref{70}). The symbol ${\rm d} [\sigma]$
denotes the flat Berezinian. Eq.~(\ref{71}) is exact.

\subsection{Saddle-Point Approximation}

Putting $\ve = 0$, ${\bf W} = 0$, ${\bf J} = 0$ we vary the exponent
in ${\cal G}_{\rm R}$ with respect to the elements of ${\bf
  \Sigma}$. That yields the saddle-point equation $\sigma (E - \lambda
\sigma) = \lambda$. At the center $E = 0$ of the GUE spectrum, the
solution of that equation is $\sigma_{s p} = - i T^{- 1}_0 {\bf L}
T_0$.  Here $T_0$ is given by
\be
T_0 = \left( \matrix{ (1 + t_{1 2} t_{2 1})^{1/2} & i t_{1 2} \cr
         - i t_{2 1} & (1 + t_{2 1} t_{1 2})^{1/2} \cr} \right) \ . 
\label{72}
\ee
The matrix $t_{1 2}$ ($t_{2 1}$) has elements $(t_{1 2})_{p s, q s'}$
($(t_{2 1})_{q s, p s'}$, respectively). The elements of $(t_{1 2},
t_{2 1})$ span the saddle-point manifold for the $(P, Q)$ correlation
function.

We use the saddle-point approximation in Eq.~(\ref{71}) and expand the
exponent up to terms of first order in $\ve$, putting ${\bf W} = 0$,
${\bf J} = 0$.  That yields the symmetry-breaking term
\be
SB_{\rm R} = {i \pi \over d} \sum_p \ve_p {\rm STr}_s \bigg(
(t_{1 2} t_{2 1})_{p p} \bigg) + {i \pi \over d} \sum_q \tilde{\ve}_q
{\rm STr}_s \bigg( (t_{2 1} t_{1 2})_{q q} \bigg) \ .
\label{73}
\ee
Similarly we put $\ve = 0$, ${\bf J} = 0$, expand the logarithm in the
exponent of Eq.~(\ref{71}), work out the traces over the level indices
$\mu$, and resum the resulting series to obtain the channel-coupling
term (present only for the open system)
\be
CC_{\rm R} = - \sum_c {\rm STr}_{p s} \ln \bigg(1 + T^{(c)} t_{1 2} t_{2 1}
\bigg) \ .
\label{74}
\ee
With the assumptions introduced above, the average $S$ matrix is
diagonal. The transmission coefficient $T^{(c)}$ in channel $c$ is
defined as $T^{(c)} = 1 - |\langle S_{c c} \rangle|^2$. As a result,
the saddle-point approximation to $\langle {\cal G}_{\rm R} \rangle$
is given by
\be
\langle {\cal G}_{\rm R} \rangle_{s p} = \int {\rm d} \mu(t)
\bigg( ... \bigg) \exp  \{ SB_{\rm R} + CC_{\rm R} \} \ . 
\label{75}
\ee
Here the dots indicate the source terms. As in Ref.~\cite{Ver85}, the
invariant measure ${\rm d} \mu(t)$ is defined by the transformation
from the variables that parametrize $\sigma$ to the ones that
parametrize $(t_{1 2}, t_{2 1})$. It turns out that it is not
neccessary to work out ${\rm d} \mu(t)$ explicitly.

\subsection{Massive Modes}

Massive modes are those degrees of freedom that do not lie in the
saddle-point manifold. As for graphs, these are treated in Gaussian
approximation. For the orthogonal case and for $P = 1 = Q$ it is shown
in Ref.~\cite{Ver85} that the massive modes lie either in the retarded
or in the advanced block. The argument carries through also in the
unitary case and for $P \geq Q \geq 1$. 

For the closed system (${\bf W} = 0$) we follow Ref.~\cite{Ver85}. We
put $\ve = 0$ and $E = 0$ for simplicity and write $\sigma = \sigma_{s
  p} + \delta \sigma$ where $\delta \sigma$ stands for the massive
modes. We expand the exponent in Eq.~(\ref{71}) in powers of $\delta
\sigma$ and keep only terms up to second order, neglecting the source
terms. With $\delta {\cal P} = T_0 \delta \sigma_{s p} T^{- 1}_0$ and
$[\delta {\cal P}, {\bf L}] = 0$ that gives in the exponent the term
\be
- N {\rm STr}_{t s} (\delta {\cal P})^2 \ .
\label{76}
\ee
In Ref.~\cite{Ver85} it is shown that for $N \to \infty$ and in the
vicinity of the saddle point, the Berezinian for $\delta {\cal P}$ is
flat. The arguments carry through also for the present case. The
source terms for the massive modes are given by the expansion of $N
\ln (1 + i {\bf L} \delta {\cal P} - i {\bf L} T_0 {\bf J} T^{- 1}_0 /
\lambda)$ in powers of ${\bf J}$. With $d = \pi \lambda / N$, the term
${\bf J} / \lambda$ is inversely proportional to $N$. The substitution
$\delta {\cal P} \to \sqrt{N} \delta {\cal P}$ then shows that all
source terms (and, therefore, the contributions of massive modes)
vanish with some inverse power of $N$. That conclusion, demonstrated
at the center $E = 0$ of the spectrum, can be shown to hold everywhere
except near the end points.

We turn to the open system. Expanding the last term in the exponent of
Eq.~(\ref{71}) in powers of ${\bf W}$ and resumming we find that for
$\ve = 0$ the term~(\ref{76}) is replaced by
\ba
&& - (N - \Lambda) {\rm STr} (\delta {\cal P})^2 + \sum_{a = 1}^\Lambda
{\rm STr}_{t s} \bigg( \frac{i {\bf L} x_a}{1 + i {\bf L} x_a
\sigma_{s p}} \delta \sigma \bigg) \nonumber \\
&& \qquad - \sum_{a = 1}^\Lambda {\rm STr}_{t s} \bigg( \sigma_{s p}
\frac{1}{1 + i {\bf L} x_a \sigma_{s p}} \delta \sigma \bigg)^2 \ .
\label{77}
\ea
Here $x_a = \pi N v^2_a / \lambda$ is of order unity (not $N$). The
sums extend over the open channels. The same substitutions as used
above, i.e., $\delta \sigma \to \delta {\cal P} \to \sqrt{N} \delta
{\cal P}$, then show that for $\Lambda$ fixed and $N \to \infty$, 
expression~(\ref{77}) reduces to expression~(\ref{76}). The source
terms for the massive modes are now given by the expansion of $N \ln
(1 + i {\bf L} \delta {\cal P} - i ({\bf L} T_0 {\bf J} T^{- 1}_0 /
\lambda) - i ({\bf L} T_0 {\bf W} T^{- 1}_0 / \lambda))$ in powers of
${\bf J}$. Eqs.~(\ref{60}) show that the matrices ${\cal S}_t$ are
dyadic products of two vectors and, thus, possess only a single
nonvanishing eigenvalue. This makes up for the fact that, in contrast
to the case of the closed system, they lack a factor $N^{- 1}$. An
expansion of the logarithmic term in powers of ${\bf W}$ embellishes
the source terms ${\bf J}$ with factors proportional to powers of
$x_a$ but does not affect the overall dependence on $N$. It follows
that the contribution of massive modes vanishes for $N \to \infty$
also for open graphs.

We have, thus, shown that the contribution of the massive modes to all
$(P, Q)$ correlation functions for the GUE vanish with some inverse
power of $N$ as $N \to \infty$. Therefore, these functions are
obtained by differentiation of $\langle {\cal G}_R \rangle_{s p}$ with
respect to the sources.

\section{Equivalence Proof}
\label{equ}

We have demonstrated that for $B \to \infty$ and $N \to \infty$ the
averaged generating functions $\langle {\cal G}_{\rm G} \rangle$ and
$\langle {\cal G}_{\rm R} \rangle$ become asymptotically equal to
$\langle {\cal G}^{(0)}_{\rm G} \rangle$ and $\langle {\cal G}_{\rm R}
\rangle_{s p}$, respectively. The equality of all $(P, Q)$ correlation
functions for quantum graphs and RMT is, thus, shown if we can prove
that
\be
\langle {\cal G}^{(0)}_{\rm G} \rangle = \langle {\cal G}_{\rm R}
\rangle_{s p} \ .
\label{78}
\ee
We show this by first constructing a one-to-one map of the RMT
saddle-point manifold unto the zero-mode manifold of graphs. We use
the transformation
\be
\tau = - i t_{1 2} \frac{1}{\sqrt{1 + t_{2 1} t_{1 2}}} \ , \
\tilde{\tau} = i t_{2 1} \frac{1}{\sqrt{1 + t_{1 2} t_{2 1}}} \ .
\label{79}
\ee
With these definitions, Eq.~(\ref{72}) and the relation $\sigma_{s p}
= - i T^{- 1}_0 {\bf L} T_0$ imply in retarded-advanced representation
\ba
\sigma_{s p} = - i \left( \matrix{
1 & \tau \cr
\tilde{\tau} & 1 \cr}
\right)
\left( \matrix{
1 & 0 \cr
0 & - 1 \cr}
\right) 
\left( \matrix{
1 & \tau \cr
\tilde{\tau} & 1 \cr}
\right)^{- 1} \ . 
\label{80}
\ea
It is shown in Ref.~\cite{Zir96} that (except for the factor $- i$
which is properly taken into account in the subsequent RMT
calculations) for a parametrization of $\sigma_{s p}$ of the
form~(\ref{80}) the integration measure is the flat Berezinian
$\prod_{p q} {\rm d} (\tilde{\tau}_{q p}, \tau_{p q})$.

The supermatrix $\tau$ ($\tilde{\tau}$) has nonzero elements $\tau_{p
  s, q s'}$ ($\tilde{\tau}_{q s, p s'}$) only in the retarded-advanced
block (in the advanced-retarded block, respectively). In each subblock
labelled $(p, q)$ that supermatrix has dimension two. The integration
measure for $(\tau, \tilde{\tau})$ is the flat Berezinian. The
Boson-Boson (Fermion-Fermion) blocks of the matrices $\tau$ and
$\tilde{\tau}$ are related by $\tilde{\tau}_{B B} = \tau^\dag_{B B}$
and by $\tilde{\tau}_{F F} = - \tau^\dag_{F F}$. The eigenvalues of
the positive definite Hermitean matrix $- \tilde{\tau}_{F F} \tau_{F
  F}$ are smaller than or equal to unity. The $BF$ blocks and the $FB$
blocks carry independent anticommuting integration variables. For each
pair of indices $(p, q)$ the pair $(\tilde{\tau}_{q p}, \tau_{p q})$
of $2 \times 2$ supermatrices possesses the same symmetry properties
as the pair $(\tilde{\tau}_{1 1}, \tau_{1 1})$ the elements of which
span the saddle-point manifold for the $(P = 1, Q = 1)$ correlation
functions. All this follows from arguments of symmetry and convergence
detailed in Ref.~\cite{Ver85} for the orthogonal case. As a
consequence of the color-flavor transformation~\cite{Zir96} all these
properties are shared by the matrices $Y$ and $\tilde{Y}$. Therefore,
there exists a one-to-one map of the two sets of matrices $(\tau_{p
  q}, \tilde{\tau}_{q p})$ and $(Y_{p q}, \tilde{Y}_{q p})$ onto each
other and we can, without loss of generality, equate these matrices,
\be
\tau_{p q} = Y_{p q} \ , \ \tilde{\tau}_{q p} = \tilde{Y}_{q p} \ . 
\label{81}
\ee
Then the RMT saddle-point manifold coincides with the zero-mode
manifold for graphs for all values of $P$ and $Q$. The
result~(\ref{81}) is not surprising since both sets of matrices
parametrize the extension of Efetov's coset space~\cite{Efe83} to the
general case of $(P, Q)$ correlation functions.

To complete the proof of Eq.~(\ref{78}) we must show that the
integrands are identical. To achieve that we identify $\ve_p / d$ with
$\kappa_p / \Delta$ for $p = 1, \ldots, P$ and $\tilde{\ve}_q / d$
with $\tilde{\kappa}_q / \Delta$ for $q = 1, \ldots, Q$. This is
necessary because the dynamics of graphs is characterized by the wave
number $k$ and that of RMT by the energy $E$. Using that and
Eqs.~(\ref{79}) and (\ref{81}) in the expression~(\ref{73}) we find
that $SB_{\rm R}$ becomes equal to $SB_{\rm G}$ as given by
Eq.~(\ref{36}). Comparing the constraints formulated in
Eqs.~(\ref{15}) and (\ref{56}) we note that the spectrum of $k$ values
is unbounded while the energy spectrum of RMT is bounded by $4
\lambda$. Both constraints can, therefore, be read as saying that the
product of the level density and the incremental wave numbers
(energies, respectively) be small compared to the length of the
spectrum. In that sense, the two sets of constraints are equivalent.

For open graphs we postulate, in addition, that the number of channels
be the same for graphs and for RMT and that $\langle S_{a a} \rangle$
and $\langle S_{\alpha \alpha} \rangle$ be pairwise equal for all
pairs $a, \alpha = 1, \ldots, \Lambda$. That implies pairwise equality
of the transmission coefficients $T^{(\alpha)}$ and $T^{(a)}$. For the
channel coupling term $CC_R$ in Eq.~(\ref{74}) that and the
above-mentioned substitutions yield the expression~(\ref{38}) for
$CC_{\rm G}$.

To demonstrate the equality of the source terms, we first address the
closed system, neglecting the incremental wave numbers and energies.
The relevant term in Eq.~(\ref{71}) is $- {\rm STr} \ln (1 + \sigma_{s
  p} {\bf J} / \lambda)$. From Eqs.~(\ref{68}) and (\ref{62}) we have
${\bf J} / \lambda = \sigma^s_3 \pi {\bf j} / N$, with ${\bf j} =
(j_+, j_-)^T$. With $\sigma_{s p}$ given by Eq.~(\ref{80}) and with
the definitions $f = ( 1- { \tilde \tau } \tau )^{ - 1}$, $\tilde f =
( 1 - \tau { \tilde \tau } )^{ - 1}$ this is written as
\ba
1 + \sigma_{s p} {\bf J} / \lambda &=& \left( \matrix{
  1 - i \sigma^s_3 \pi j_+ / N & 0 \cr
  0 & 1 + i \sigma^s_3 \pi j_- / N \cr} \right)
 \nonumber \\
&& + 2 i \left( \matrix{
  - \tau \tilde{\tau} \tilde{f} \sigma^s_3 \pi j_+ / N &
    \tau f \sigma^s_3 \pi j_- / N \cr
  - \tilde{\tau} \tilde{f} \sigma^s_3 \pi j_+ / N  &
    \tilde{\tau} \tau f \sigma^s_3 \pi j_- / N \cr} \right) \ .
\label{82}
\ea
Denoting the first matrix on the right-hand side by $M$ we multiply
Eq.~(\ref{82}) on the right by $M^{- 1}$. We take account only of
terms linear in the source terms. Therefore multiplication of the
second matrix with $M^{- 1}$ leaves that matrix unchanged. Moreover,
we have $\ln [(1 + \sigma_{s p} {\bf J} / \lambda) M^{- 1}] = \ln(1 +
\sigma_{s p} {\bf J} / \lambda) - \ln M$. Since $M$ does not contain
any integration variables it does not contribute to the connected part
of the correlation functions. Actually, $M$ represents the
contributions due to the average level density mentioned below
Eq.~(\ref{58}).  Indeed, after differentiation with respect to $j_p$
and $j_q$ that matrix contributes the expected factors $\pi$. For the
comparison with the result for graphs (which accounts only for the
fluctuating part of the level density) we omit $M$. All this is
equivalent to replacing the matrix $M$ in Eq.~(\ref{82}) by the unit
matrix. For arbitrary supermatrices $a, b, c, d$ we use the identity
\be
{\rm STr}_{t s} \ln \left( \matrix{
a & b \cr
c & d \cr}
\right) = {\rm STr}_{t s} \ln ( a - b d^{- 1} c) + {\rm STr}_{t s} \ln d
\ .
\label{83}
\ee
With $d = 1 + 2 i \tilde{\tau} \tau f \sigma^s_3 \pi j_- / N$ and
$\tilde{d} = 1 + 2 i \tau \tilde{f} \sigma^s_3 \tilde{\tau}\pi j_- /
N$ we have $\ln d = \ln \tilde{d}$. Thus, expression~(\ref{83}) equals
$\ln ( \tilde{d} (a - b d^{- 1} c) )$. Using this we obtain
\ba && 
- {\rm STr} \ln (1 + \sigma_{s p} {\bf J} / \lambda) = - {\rm
  STr} \ln \bigg( \bigg[ \big( 1 + 2 i \tau f \sigma^s_3 (\pi j_- / N)
  \tilde{\tau} \big) \nonumber \\
&& \qquad \times \big( 1 - 2 i \tau \tilde{\tau} \tilde{f}
  \sigma^s_3 (\pi j_+ / N) \big) \bigg] - 4 \bigg[ \big( 1 + 2 i \tau
  f \sigma^s_3 (\pi j_- / N) \tilde{\tau} \big) \nonumber \\
&& \qquad \times \tau f \sigma^s_3 (\pi j_- / N) \frac{1}{1 + 2 i
  \tilde{\tau} \tau f \sigma^s_3 (\pi j_- / N)} \tilde{\tau} \tilde{f}
  \sigma^s_3 (\pi j_+ / N) \bigg] \bigg) \ .
\label{84}
\ea
We insert in the last line of this expression behind the first factor
$j_-$ the identity, written as $\tilde{\tau} (\tilde{\tau})^{-
  1}$. Then the first factor in round brackets commutes with $\tau f
\sigma^s_3 \tilde{\tau}$. We use
\be
\big( 1 + 2 i \tau f \sigma^s_3 (\pi j_- / N) \tilde{\tau} \big)
(\tilde{\tau})^{- 1} \frac{1}{1 + 2 i \tilde{\tau} \tau f
\sigma^{\rm s}_3 (\pi j_- / N)} = (\tilde{\tau})^{- 1}
\label{85}
\ee
and obtain
\ba
&& - {\rm STr} \ln (1 + \sigma_{s p} {\bf J} / \lambda) = {\rm STr}
\ln ( 1 - \tau \tilde{\tau}) \nonumber \\
&& - {\rm STr} \ln \bigg( 1 - (1 + 2 i \pi \sigma^s_3 (j_+ / N) )
\tau (1 - 2 i \pi \sigma^s_3 (j_- / N)) \tilde{\tau} \bigg) \ .
\label{86}
\ea
Here and in Eqs.~(\ref{87}) and (\ref{88}), the first term on the
right-hand side is obviously not a source term. With the
identification~(\ref{81}) the right-hand side becomes
\be
{\rm STr} \ln ( 1 - Y \tilde{Y}) - {\rm STr} \ln \bigg( 1 - (1 +
2 i \pi \sigma^s_3 j_+ / N ) Y (1 - 2 i \pi \sigma^s_3 j_- / N)
\tilde{Y} \bigg) \ .
\label{87}
\ee
For graphs, we use Eqs.~(\ref{35}), $z_+ = 1 = z_-$, the implicit
definitions of the matrices ${\cal B}$ in Eqs.~(\ref{25}), and
Eqs.~(\ref{20}). We also use that $[\sigma^d_1 \Sigma^{(B)}, Y] = 0$,
and we replace $L_b$ by the average value $\sum_b L_b / B$. That
gives
\be
{\rm STr} \ln ( 1 - Y \tilde{Y}) - {\rm STr} \ln \bigg( 1 - (1 +
2 i \pi \sigma^s_3 j_+ / (2 B) ) Y (1 - 2 i \pi \sigma^s_3 j_- / (2 B))
\tilde{Y} \bigg) \ .
\label{88}
\ee
The factors $1 / N$ and $1 / (2 B)$ are the inverses of the dimensions
$N$ and $2 B$ of the matrices $H$ and $\Sigma^{(B)}$, respectively.
These are sent to infinity after the differentiations in
Eqs.~(\ref{23}) and (\ref{64}) are carried out. Without loss of
generality we may, therefore, put $N = 2 B$. Then the source
terms~(\ref{87}) and (\ref{88}) are identical.

For open systems, the matrices ${\bf W}$ and $\sigma^d_1 \Sigma^{(B)}$
are not related in any obvious way. Therefore, the equality of the
source terms can be demonstrated only after the terms involving these
matrices have been converted into terms involving the average $S$
matrix and/or the transmission coefficients. For the random-matrix
approach we use the last term in Eq.~(\ref{71}) taken at the saddle
point, putting $\ve = 0$ and $E = 0$. In the calculations that follow
we repeatedly suppress terms of order zero in ${\bf J}$ without
mention as these are fully taken into account by the channel-channel
coupling term~(\ref{74}). In the term
\be
- {\rm STr} \ln \bigg( 1 + \frac{1}{1 +  i \lambda^{- 1} \sigma_{s p}
{\bf W}} \lambda^{- 1} \sigma_{s p} {\bf J} \bigg) 
\label{89}
\ee
we expand the denominator in powers of ${\bf W}$, use $(W W^\dag)_{a
  b} = \delta_{a b} N v^2_a$, and resum the result. With $x =
\{\delta_{a b} \pi^2 v^2_a / d \}$ that gives
\be
- {\rm STr} \ln \bigg( 1 + \frac{1}{1 +  i x \sigma_{s p} {\bf L}}
\lambda^{- 1} \sigma_{s p} {\bf J} \bigg) \ .
\label{90}
\ee
We expand the logarithm in powers of ${\bf J}$, use the
definitions~(\ref{68}), (\ref{62}), and (\ref{60}), perform the trace
over the level index $\mu$ and resum the series. We define in channel
space (not in level space) the matrix ${\bf j}$ with elements $j_p
\delta_{a a_p} \delta_{b b_p}$ in the retarded sector and $j_q \delta_{a
\tilde{b}_q} \delta_{b \tilde{a}_q}$ in the advanced sector. That gives
\be
- {\rm STr}_{a t s} \ln \bigg( 1 - \frac{1}{1 +  i x \sigma_{s p}
{\bf L}} 2 i x \sigma_{s p} {\bf L} \sigma^s_3 {\bf j} \bigg) \ .
\label{91}
\ee
We write this as $- {\rm STr} \ln (1 + i x \sigma_{s p} {\bf L} - 2 i
x \sigma_{s p} {\bf L} \sigma^s_3 {\bf j})$ with the convention that
the channel index on $x$ is determined by the closest factor ${\bf j}$
to the right of $x$. We follow the steps that lead from Eq.~(\ref{82})
to Eq.~(\ref{87}). Effectively this amounts in Eq.~(\ref{86}) to the
replacements $i \pi \sigma^s_3 j_+ / N \to [ - x / ( 1 + x)] + 2
\sigma^s_3 j_+ x / ( 1 + x)^2$ and $- i \pi \sigma^s_3 j_- / N \to [-
  x / ( 1 + x)] + 2 \sigma^s_3 j_- x / ( 1 + x)^2$. For fixed $a$ the
resulting terms $1 - 2 x_a / (1 + x_a)$ are equal to the elements
$\langle S_{a a} \rangle$ of the average $S$ matrix. With the
transmission coefficients given by $T^{(a)} = 4 x_a / (1 + x_a)^2$ and
with $\tau \to Y$, $\tilde{\tau} \to \tilde{Y}$ we thus obtain
\be
- {\rm STr}_{a t s} \ln \bigg( 1 - (\langle S \rangle + T \sigma^s_3
{\bf j}_+) Y (\langle S \rangle + T \sigma^s_3 {\bf j}_- ) \tilde{Y}
\bigg) \ .
\label{92}
\ee
We have defined ${\bf j}_+ = j_+ \delta_{a a_p} \delta_{b b_p}$ and
correspondingly for ${\bf j}_-$. The matrices $\langle S \rangle $ and
$T$ are diagonal in channel space with elements $\langle S_{a a}
\rangle$ and $T^{(a)}$, respectively. The analogue of the matrix $M$
(first term on the right-hand side of Eq.~(\ref{82})) is now given by
$(1 + (\langle S \rangle - 1) \sigma^s_3 {\bf j})$. After
differentiation with respect to $j_p$ and $j_q$, that matrix
contributes terms of the form $\langle S \rangle - 1$ to the
$S$-matrix correlators. Such terms must arise because the $S$-matrix
correlators for the RMT case are by construction (see Eq.~(\ref{61})
and the remark below Eq.~(\ref{58})) averages over products of
elements of $(S - 1)$. With $S - 1 = (\langle S \rangle - 1) + S^{\rm
  fl}$ they contain factors $(\langle S \rangle - 1)$. We disregard
these contributions and, thus, the analogue of $M$ because we aim at
comparing the source terms for $S^{\rm fl}$ for RMT and for graphs.

For open graphs we use Eq.~(\ref{35}), the implicit
definitions~(\ref{25}) for the matrices ${\cal B}$, and the
definitions~(\ref{20}). The source terms have the form
\be
- {\rm STr} \ln \bigg( 1 - \big( \sigma^d_1 \Sigma^{(B)} - \sigma^s_3
j_+ {\cal I}_+ \big) Y  \big( (\Sigma^{(B)})^\dag \sigma^d_1 - \sigma^s_3
j_- {\cal I}^\dag_- \big) \tilde{Y} \bigg) \ .
\label{93}
\ee
Since $\sigma^d_1$ commutes with $\sigma^s_3 j_\pm$ and $Y$,
  $\tilde{Y}$ and since $(\sigma^d_1)^2 = 1$ this is equal to
\be
- {\rm STr} \ln \bigg( 1 - \big( \Sigma^{(B)} - \sigma^s_3
j_+ {\cal S}_+ \big) Y  \big( (\Sigma^{(B)})^\dag - \sigma^s_3
j_- {\cal S}^\dag_- \big) \tilde{Y} \bigg) \ .
\label{94}
\ee
We go to the vertex representation, $\Sigma^{(B)} \to \Sigma^{(V)}$
and the associated changes $S \to S^{(V)}$, $\tilde{S} \to
\tilde{S}^{(V)}$. This is permissible because $j_\pm$, $\sigma^s_3$,
$Y$, and $\tilde{Y}$ do not depend on directed bond indices. We recall
that $\Sigma^{(V)}$ is block diagonal, the matrices
$\sigma^{(\alpha)}$ occupying the diagonal blocks. For $\alpha >
\Lambda$ these are unitary, and the source terms in these blocks
vanish. That leaves us with
\be
- {\rm STr}_{\alpha s t} \ln \bigg( 1 - \big( \Sigma^{(B)} - \sigma^s_3
j_+ {\cal S}_+ \big) Y  \big( (\Sigma^{(B)})^\dag - \sigma^s_3 j_-
{\cal S}^\dag_- \big) \tilde{Y} \bigg) \ .
\label{95}
\ee
The trace extends only over the blocks with $\alpha \leq \Lambda$. In
each such block we diagonalize $\sigma \to U_1 \sigma_{\rm diag} U_2$,
with unitary matrices $U_1, U_2$. As shown in Ref.~\cite{Plu13a}, all
diagonal elements but the first one of $\sigma^{(\alpha)}_{\rm diag}$
have magnitude unity, the first one being given by $- \rho^{(\alpha)}$
(we suppress here a phase factor which cancels anyway). Each of the
source terms $S^{(V)}$ and $\tilde{S}^{(V)}$ is the dyadic product of
two vectors. Upon transforming $S^{(V)} \to U^\dag_1 S^{(V)} U^\dag_2$
and $\tilde{S}^{(V)} \to U_2 \tilde{S}^{(V)} U_1$, each of these
vectors has a single nonvanishing element only. As a result all
elements of the transformed matrices $U^\dag_1 S^{(V)} U^\dag_2$ and
$U_2 \tilde{S}^{(V)} U_1$ but the $(1, 1)$ element vanish. The latter
has the value $(T^{(\alpha_p)})^{1 / 2} (T^{(\beta_p)})^{1 / 2} = T^{1
  / 2} \delta_{\alpha \alpha_p} \delta_{\beta \beta_p} T^{1 / 2}$ in
the retarded space and $(T^{(\delta_q)})^{1 / 2} (T^{(\gamma_q)})^{1 /
  2} = T^{1 / 2} \delta_{\alpha \delta_q} \delta_{\beta \gamma_q} T^{1
  / 2}$ in the advanced space. Here $T$ is a diagonal matrix in
channel space with elements $T^{(\alpha)}$. With ${\bf j}_+ = j_+
\delta_{\alpha \alpha_p} \delta_{\beta \beta_p}$ and ${\bf j}_- = j_-
\delta_{\alpha \delta_q} \delta_{\beta \gamma_q}$
expression~(\ref{95}) becomes
\be
- {\rm STr}_{\alpha s t} \ln \bigg( 1 - \big( \langle S \rangle + T^{1 / 2}
\sigma^s_3 {\bf j}_+ T^{1 / 2} \big) Y  \big( \langle S \rangle + T^{1 / 2}
\sigma^s_3 {\bf j}_- T^{1 / 2} \big) \tilde{Y} \bigg) \ .
\label{96}
\ee
Expanding this into a Taylor series, using in each term that $T$
commutes with $Y$ and $\langle S \rangle$, using the cyclic invariance
of the trace, and resumming we obtain
\be
- {\rm STr}_{\alpha s t} \ln \bigg( 1 - \big( \langle S \rangle + T
\sigma^s_3 {\bf j}_+ \big) Y  \big( \langle S \rangle + T \sigma^s_3
{\bf j}_- \big) \tilde{Y} \bigg) \ ,
\label{97}
\ee
in perfect agreement with expression~(\ref{92}). This completes the
proof of Eq.~(\ref{78}).

\section{Summary and Conclusions}
\label{con}

The combination of the results derived in the present paper with those
obtained in Refs.~\cite{Plu14, Plu13b} amounts to a complete proof of
the BGS conjecture for the cases of orthogonal and unitary symmetry.
We have not considered the symplectic case. The proof holds in the
limit $B \to \infty$ of infinite graph size for simple connected
graphs with the following three properties. (i) The bond lengths must
be incommensurate. (ii) The eigenvalue $+ 1$ of the Perron-Frobenius
operator must be separated by a finite gap from the rest of the
spectrum. (iii) For all $(\alpha, \beta)$ the elements of the vertex
scattering matrices must obey $|\sigma^{(\alpha)}_{\beta \beta}
\sigma^{(\beta)}_{\alpha \alpha}| \leq b < 1$.

The proof given in the present paper is tailored to the unitary
case. It involves the following steps. (a) The $(P, Q)$ correlation
functions for levels and $S$-matrix elements are witten as derivatives
of generating functions. These are expressed as superintegrals. The
average over wave number $k$ is performed with the help of property
(i) and of the color-flavor transformation. These steps are exact and
yield Eqs.~(\ref{31}) and (\ref{32}) for the averaged generating
functions. (b) Expansion of the effective action in Eq.~(\ref{32}) up
to terms of second order in the integration variables and
transformation to the eigenvector representation of the
Perron-Frobenius operator yields the bilinear form~(\ref{38}). In that
form the mode corresponding to eigenvalue $\lambda_1 = 1$ is
absent. That mode is identified as the zero mode $(Y, \tilde{Y})$.
(c) The contribution of the remaining modes is calculated using the
loop expansion. Under conditions (ii) and (iii) the contributions of
the resulting Gaussian superintegrals to every $(P, Q)$ correlation
function vanish for $B \to \infty$. In that limit, the generating
functions are entirely determined by the zero mode. (iv) The
equivalence with RMT is established in terms of a one-to-one map of
the zero-mode manifold for graphs onto the saddle-point manifold for
RMT, and of an identification of all terms in the generating
functions. With a slight change of notation and under corresponding
conditions on the vertex matrices, our proof carries over to the
orthogonal case. Our results hold asymptotically as we have only shown
that every contribution of the massive modes vanishes individually in
the limit $B \to \infty$. Therefore, we cannot calculate leading-order
corrections in $1 / B$.

In contrast to previous approaches~\cite{Gnu04, Gnu05, Plu13a, Plu13b,
  Plu14} we have not used the saddle-point approximation. Direct use
of the zero mode makes the proof stringent, simple, and transparent.
We have removed the restriction to completely connected graphs used
earlier~\cite{Plu13a, Plu13b, Plu14}. In view of earlier work on the
two-point function~\cite{Gnu04, Gnu05, Plu13a, Plu13b} our result is
probably expected although it would seem conceivable that differences
between graphs and RMT might have existed which, while absent for the
two-point functions, would systematically increase with increasing $P$
and $Q$. We have shown that such differences do not exist.

We believe that our derivation and result are of general interest. The
Perron-Frobenius operator is bistochastic and has a single eigenvalue
$\lambda_1 = 1$. We have shown that this fact directly implies the
existence of the zero mode. The zero-mode manifold represents the
generalization to arbitrary values of $P$ and $Q$ of Efetov's coset
spaces~\cite{Efe83} for unitary or orthogonal symmetry. Aside from the
incremental wave numbers and, for open graphs, from the strengths of
the couplings to the open channels, the zero-mode part~(\ref{46}) of
the generating functions carries no information on the graph actually
considered. The zero mode is universal and entirely governed by
symmetry. These results are relevant in the limit $B \to \infty$ where
the contributions of the massive modes disappear with inverse powers
of $B$, and where the generating functions for graphs and for RMT
coincide. That agreement validates the BGS conjecture for graphs.

We discuss our three essential assumptions. (i) Incommensurability of
the bond lengths $L_b$ guarantees ergodicity and makes it possible to
calculate the average over wave number $k$ in terms of an average over
independent phases $\phi_b = k L_b$. That allows us to show that the
BGS conjecture holds for every individual graph with incommensurate
bond lengths. Alternatively we might consider an ensemble of graphs
with a distribution of statistically independent bond lengths $L_b$.
Averaging over that distribution would likewise yield an average over
independent phases $\phi_b$. Thus, the BGS conjecture holds on average
for an ensemble of graphs irrespective of any condition on bond
lengths. Therefore, incommensurability of the bond lengths $L_b$ is a
less stringent requirement than our conditions (ii) and (iii). For
individual graphs with bond lengths that are partly commensurate we
do, of course, expect deviations from RMT fluctuation properties. (ii)
Classically, the existence of a gap in the spectrum of the
Perron-Frobenius operator guarantees that the graph is mixing. In the
quantum context condition (ii) ensures that contributions of the modes
$(z_j, \tilde{z}_j)$ to the loop expansion vanish asymptotically.
While sufficient, condition (ii) may not be necessary for the proof of
the BGS conjecture. A weaker condition (power-law
suppression~\cite{Tan01, Gnu05,Gnu08, Gnu10} of the eigenvalue density
near $\lambda_1 = 1$) might suffice to guarantee classical chaos
(albeit not mixing), uniform spreading of eigenfunctions over the
graph, and even disappearance of the contributions of the massive
modes for $B \to \infty$ provided the loop expansion could be
technically avoided. We have not addressed these questions. (iii) For
all $\alpha, \beta$ the diagonal elements of the vertex scattering
matrices obey $|\sigma^{(\alpha)}_{\beta \beta} \sigma^{(\beta)}_{\alpha
  \alpha}| \leq b < 1$. That assumption excludes the formation of
bound states on single bonds that are completely separated from the
rest of the graph. Such states might affect the spectral fluctuation
properties.

The conditions (ii) and (iii) are physically plausible, and deviations
from universal behavior are expected if one is violated. Take
condition (ii), for instance, and let us consider two completely
connected simple chaotic graphs $g_1$ and $g_2$ that are connected
with each other by a single bond. In the limit of infinite graph size
we expect that the single bond plays an ever diminishing role, and
that the spectral properties are dominated by the zero-mode manifolds
for $g_1$ and $g_2$. We conjecture that the spectrum is a
superposition of two GOE (or two GUE) spectra with weak repulsion
between levels from $g_1$ and those from $g_2$. Another case is a
chain of completely connected simple chaotic graphs with few bonds
connecting only neighboring members in the chain. Here Anderson
localization might be expected, causing strong deviations from
Wigner-Dyson statistics.

Are there graphs that satisfy conditions (ii) and (iii) and if so, how
numerous are they? As for the first part of the question, there are
examples in the orthogonal case where the spectrum of the
Perron-Frobenius operator possesses a gap~\cite{Har07}. If
time-reversal invariance is violated by a magnetic field that affects
the phases of the bond propagators but not the matrices
$\sigma^{(\alpha)}$, the spectrum of the Perron-Frobenius operator is
unaffected and the gap persists in the unitary case. Concerning
condition (iii), bound states on single bonds with total reflection at
the adjacent vertices require very special boundary
conditions. Therefore, we conjecture that condition (iii) is satisfied
for most Hermitean boundary conditions imposed at the vertices. The
second part of the question requires a deeper understanding of the
relationship between the boundary conditions at the vertices and the
spectrum of the Perron-Frobenius operator~\cite{Smi96, Kot99, Ber01,
  Tan01}. We consider this an important challenge for future
investigations.

The symmetry-breaking terms for graphs (for RMT) are obtained by
expanding the effective action in powers of the wave-number increments
$\kappa_p, \tilde{\kappa}_q$ (the energy increments $\ve_p,
\tilde{\ve}_q$, respectively). With the identifications defined in
Section~\ref{equ} both terms agree to lowest order in these
increments. Inspection shows that the terms of next order do not
agree. Therefore, the agreement of the fluctuation properties of
graphs with those of RMT is limited to a wave-number interval defined
by the range of validity of the approximations used. With $\kappa$
representing any of the $\kappa_p$ or $\tilde{\kappa}_q$, we have
expanded $\exp \{ i \kappa L_b \}$ in a Taylor series and kept terms
of order zero and one only. The neglect of higher-order terms is
justified if $\kappa L_b \ll 1$ for all $b$ or, equivalently, if
$\kappa \ll 1 / L_{\rm max}$. With $\overline{L}$ the average bond
length, the dimensionless variable $\kappa / \Delta$ must, therefore,
obey $\kappa / \Delta \ll B \overline{L} / L_{\rm max}$. That bound
tends to infinity with $B$, and so does the range of agreement of the
$(P, Q)$ correlation functions for graphs and for RMT. This situation
differs from the case of the semiclassical approximation where the
spectral fluctation properties agree with RMT predictions in an energy
interval defined by the shortest periodic orbit.

In Refs.~\cite{Aga95, And96a, And96b} use of the classical PF operator
was advocated as a sufficient means to characterize the spectral
fluctuation properties of chaotic quantum systems. For the modes
$(z_j, \tilde{z}_j)$ (Eq.~(\ref{38})) the present work confirms that
suggestion: Their masses are determined by the eigenvalues of the PF
operator. However, knowledge of the PF operator is not sufficient for
the non-diagonal massive modes. Their masses depend on the matrix
elements of the vertex scattering matrix $\Sigma^{(B)}$, a quantum
operator, see Eq.~(\ref{38}).

{\it Acknowledgements}. The authors are grateful to E. Bogomolny,
P. Cejnar, S. Gnutzmann, P. LeBoeuf, P. Kurasov, J. Kvasil, and
U. Smilansky for useful comments. Thanks are due to an anonymous
referee for helpful and constructive criticism. ZP acknowledges
  support by the Czech Science Foundation under Project No P203 - 13 -
  07117S.

\end{document}